\documentclass[journal]{IEEEtran}
\ifCLASSINFOpdf
\else
\fi
\usepackage[utf8]{inputenc}
\usepackage{amsmath,amssymb,amsfonts}
\usepackage{cite}

\usepackage{algorithmic}
\usepackage{stfloats}
\usepackage{graphicx}
\usepackage{textcomp}
\usepackage{xcolor}
\usepackage{color}
\usepackage{pifont}
\usepackage{amsthm}

\usepackage{hyperref}
\usepackage{mathrsfs}
\usepackage{booktabs}
\usepackage{hyperref}
\usepackage{cases}
\usepackage{comment}
\usepackage{empheq} 
\usepackage{caption}
\usepackage{subcaption}
\usepackage{extarrows}
\allowdisplaybreaks[4]

\usepackage{url}  
\usepackage{graphicx}  
\usepackage[ruled,vlined,linesnumbered]{algorithm2e}

\usepackage{listings}

\begin{document}
\captionsetup[figure]{labelfont={rm},labelformat={default},labelsep=period,name={Fig.}}
\title{Robust Optimization for 
Movable Antenna-aided Cell-Free ISAC with Time Synchronization Errors}
%
%

\author{Yue Xiu,~\IEEEmembership{Member,~IEEE},~Yang Zhao,~\IEEEmembership{Member,~IEEE},~Ran Yang,~Wanting Lyu,~\IEEEmembership{Member,~IEEE},\\Dusit Niyato,~\IEEEmembership{Fellow,~IEEE},~Dong In Kim,~\IEEEmembership{Life Fellow,~IEEE},\\~Guangyi Liu,~\IEEEmembership{Senior Member,~IEEE},~Ning Wei,~\IEEEmembership{Member,~IEEE}
\thanks{Yue Xiu, Ran Yang, Wanting Lyu and Ning Wei are with National Key Laboratory of Science and Technology on Communications, University of Electronic Science and Technology of China, Chengdu 611731, China (E-mail: xiuyue12345678@163.com, yangran6710@outlook.com, lyuwanting@yeah.net, wn@uestc.edu.cn).}
\thanks{Yang Zhao and Dusit Niyato are with the College of Computing and
Data Science, Nanyang Technological University, Singapore, 639798 (e-mail:zhao0466@e.ntu.edu.sg, dniyato@ntu.edu.sg).}
\thanks{Dong In Kim is with the Department of Electrical and Compute Engineering, Sungkyunkwan University, Suwon 16419, South Korea (e-mail: dongin@skku.edu).}
\thanks{G. Liu is with the China Mobile Research Institute, Beijing 100053, China
(email: liuguangyi@chinamobile.com).}

}

\maketitle
\begin{abstract}

The cell-free integrated sensing and communication (CF-ISAC) system, which effectively mitigates intra-cell interference and provides precise sensing accuracy, is a promising technology for future 6G networks. However, to fully capitalize on the potential of CF-ISAC, accurate time synchronization (TS) between access points (APs) is critical. Due to the limitations of current synchronization technologies, TS errors have become a significant challenge in the development of the CF-ISAC system. In this paper, we propose a novel CF-ISAC architecture based on movable antennas (MAs), which exploits spatial diversity to enhance communication rates, maintain sensing accuracy, and reduce the impact of TS errors. We formulate a worst-case sensing accuracy optimization problem for TS errors to address this challenge, deriving the worst-case Cramér-Rao lower bound (CRLB). Subsequently, we develop a joint optimization framework for AP beamforming and MA positions to satisfy communication rate constraints while improving sensing accuracy. A robust optimization framework is designed for the highly complex and non-convex problem. Specifically, we employ manifold optimization (MO) to solve the worst-case sensing accuracy optimization problem. Then, we propose an MA-enabled meta-reinforcement learning (MA-MetaRL) to design optimization variables while satisfying constraints on MA positions, communication rate, and transmit power, thereby improving sensing accuracy. The simulation results demonstrate that the proposed robust optimization algorithm significantly improves the accuracy of the detection and is strong against TS errors. Moreover, compared to conventional fixed position antenna (FPA) technologies, the proposed MA-aided CF-ISAC architecture achieves higher system capacity, thus validating its effectiveness.

\end{abstract}


\section{Introduction} Integrated sensing and communication (ISAC) technology has emerged as a key application in 6G systems, including smart cities, autonomous driving, and remote healthcare\cite{10707080,10666854}. ISAC integrates communication and sensing functions into a shared hardware platform, utilizing common spectrum resources to improve spectrum and hardware efficiency\cite{9598915}. Meanwhile, cell-free (CF) systems, an emerging architecture, are gaining attention for their advantages in distributed access points (APs) collaboration. In CF systems, multiple APs are distributed over a wide area and connected to a central processing unit (CPU) via a backhaul network, providing services to user equipment (UEs) while mitigating interference caused by cell boundaries in traditional cellular networks. Therefore, combining ISAC with CF not only enhances interference resistance and network coverage balance, but also improves communication reliability in high-speed mobile environments. Consequently, cell-free integrated sensing and communication (CF-ISAC) systems will become a critical technology for future 6G wireless communication networks.

\textbf{Challenges.} In practical applications, time synchronization (TS) of APs in CF-ISAC systems is often inaccurate due to system limitations, leading to degraded sensing performance. Precise TS is crucial to fully exploit the benefits of cooperative sensing in CF-ISAC systems. However, although existing TS errors compensation schemes can achieve satisfactory performance in communication systems, they inevitably introduce ambiguity in time delay estimation during the sensing phase, thereby degrading the overall sensing accuracy of the CF-ISAC system.

\textbf{Motivation.} To further enhance the anti-jamming capability and robustness against TS errors in CF-ISAC systems, increasing the number of distributed APs can be effective, but requires high power consumption and costs. In addition,  emerging reconfigurable intelligent surface (RIS) technology can improve CF-ISAC performance through passive components, but its deployment is constrained by environmental elements. Fortunately, a promising technology, movable antennas (MAs), provides an energy-efficient alternative to enhance CF-ISAC systems' capacity and robustness. Unlike traditional fixed-position antennas (FPAs), MAs offer the flexibility of dynamic antenna movement\cite{10709885}, enabling adequate compensation for TS errors in dynamic environments. MA technology adjusts antenna positions and orientations in real time based on TS errors, reducing signal distortion. This flexibility allows the system to suppress interference caused by TS errors, optimize signal transmission, and significantly improve the communication rate and sensing accuracy. Therefore, MA technology is considered an effective method in mitigating the degradation of system performance caused by TS errors.

To address the impact of TS errors in CF-ISAC systems, we propose a CF-ISAC architecture aided by MAs to enhance robustness against TS errors and improve overall system performance. However, this problem faces several challenges, especially in evaluating the effects of TS errors on sensing performance and enhancing the system's robustness to TS errors. To overcome these challenges, this paper proposes an MA-aided CF-ISAC system, where multiple APs coordinate beamforming transmissions and MA positions to simultaneously support user communications and enable static target sensing. The main contributions of this work are summarized as follows.
\begin{itemize}
    \item First, the Cramér–Rao lower bound (CRLB) for target position estimation is derived considering the influence of TS errors, highlighting the inherent coupling between TS errors and sensing accuracy. Given that TS errors are bounded within known intervals but cannot be precisely specified, a worst-case analysis is conducted to obtain the corresponding worst-case CRLB, thereby improving robustness against TS errors uncertainty. Specifically, this worst-case CRLB is then used as the objective function in an optimization problem, where TS errors are optimized. As the resulting problem is non-convex, a manifold optimization (MO) algorithm is employed to determine the TS errors. 
    \item Given the specified TS errors, the problem is formulated as a worst-case CRLB minimization problem, jointly optimizing transmit beamforming, MA positions, and subject to constraints on communication QoS, transmit power, and antenna positions. To resolve this problem, an MA-enabled meta-reinforcement learning (MA-MetaRL) is proposed for the joint design of transmit beamforming and MA positions. Compared with conventional reinforcement learning (RL) approaches, the proposed MA-MetaRL algorithm enables rapid adaptation to varying user scenarios and constraints via fast policy transfer across different environments. Regarding computational complexity, the MA-MetaRL algorithm integrates meta-learning principles to significantly reduce training time while improving the policy's generalization capability. To further reduce computational complexity, in the MA-MetaRL algorithm, we introduce a twin delayed deep deterministic policy gradient (TD3) algorithm, which aids MetaRL in efficiently solving the local optimal beamforming and MA positions under multiple constraints.
    \item The simulation results demonstrate that the MA significantly improves the sensing accuracy of the CF-ISAC system while meeting the communication rate constraints and improving the robustness to TS errors. In particular, when TS errors are unknown, the proposed optimization framework, based on MO and MA-MetaRL algorithms, effectively improves robustness to TS errors while minimizing performance degradation. 
\end{itemize}

\section{Related Work}

\subsection{Resource Allocation Over CF-ISAC}
Extensive research has been devoted to resource allocation strategies in CF-ISAC systems, aiming to effectively balance communication and sensing performance. In\cite{10516289}, the authors presented a systematic analysis of the communication–sensing coverage region and proposed a dynamically adjustable resource allocation strategy to enhance resource utilization and sensing reliability in multi-user scenarios. In\cite{10494224}, a joint target detection and power allocation algorithm was proposed for multi-APs cooperation, targeting at minimize sensing errors while ensuring communication transmission rate.
To address potential security threats, \cite{10605793} studied secure resource allocation strategies for CF-ISAC systems in the presence of communication and sensing eavesdroppers, and a joint optimization framework was proposed to simultaneously enhance the secrecy rate and sensing accuracy.
In \cite{10742632}, the integration of non-orthogonal multiple access (NOMA) into CF-ISAC systems was investigated, where a joint user pairing and beamforming design was proposed to significantly enhance both system capacity and sensing accuracy.
In the context of user-centric architectures, \cite{10207026} investigated resource allocation from a user-oriented perspective. By jointly optimizing user scheduling, power allocation, and beamforming design, the proposed approach significantly enhanced overall resource utilization.
Considering both communication security and sensing quality, \cite{10540103} proposed a multi-objective resource allocation scheme that jointly maximizes the secrecy rate and the sensing signal-to-noise ratio (SNR).
Furthermore, \cite{10540103} incorporated RIS and full duplex transmission into an integrated CF-ISAC framework, further expanding degrees of freedom and the optimization potential in resource allocation. In summary, existing studies on resource allocation in CF-ISAC systems focus primarily on balancing the trade-off between communication and sensing, optimizing security-aware designs, enabling efficient multi-user scheduling, and beamforming optimization strategies.

\subsection{CF-ISAC System With TS Errors}

With the rapid advancement of ISAC technologies, CF-ISAC systems have emerged as a promising solution owing to their distributed architecture, enhanced reliability, and extensive coverage. %
However, TS errors among distributed APs pose a significant challenge to maintaining reliable communication and accurate sensing. Enhancing the robustness of CF-ISAC systems against such errors has therefore become a critical challenge in waveform design and resource allocation. Recent research has investigated various strategies to mitigate the adverse effects of TS errors in CF-ISAC architectures. %
In \cite{10834811}, a low-complexity sensing-aided TS algorithm was proposed for unmanned aerial vehicle (UAV)-enabled orthogonal frequency-division multiplexing (OFDM) systems. By leveraging sensing information, the proposed method reduced reliance on precise synchronization and enhanced the overall robustness of the system. %
In \cite{10684491}, the authors investigated coordinated transmit beamforming in a networked ISAC environment, taking into account the joint impact of channel state information (CSI) imperfections and TS errors. 
A unified beamforming optimization framework was developed, which achieved enhanced sensing accuracy while satisfying communication QoS constraints. %
To further address the effects of TS errors, \cite{10649809} formulated a joint beamforming and frame structure optimization problem. %
The proposed method balanced communication and sensing performance under various synchronization conditions by introducing guard intervals and employing a robust design based on error statistics.

\subsection{MA-Aided ISAC System}

In recent years, the rapid advancement of ISAC technologies has revealed the limitations of conventional FPA systems in dynamic environments, especially in terms of sensing accuracy and communication reliability. %
To address these challenges, the integration of MAs into ISAC frameworks has been proposed, enabling improved adaptability and performance in dynamic environments. Recent studies have focused on various aspects of MA-aided ISAC, including theoretical modeling, algorithm design, and system implementation, thereby forming a well-structured foundation for further development. %
Specifically, \cite{10599127} provided a comprehensive survey of AI-enabled MA-aided ISAC systems, highlighting key challenges and outlining future research directions, including antenna position control, dynamic channel modeling, and intelligent resource management. %
In \cite{10839251}, the authors proposed to exploit the degrees of freedom (DoF) offered by MA positions to enhance both sensing and communication performance, demonstrating the practical advantages of MAs in ISAC systems. %
Building on this, \cite{10696953} formulated a joint optimization framework for multi-user MA-ISAC systems to minimize the CRLB and developed an efficient resource allocation scheme that balanced sensing accuracy and communication quality. %
For aerial platforms, \cite{10870338} proposed a real-time beamforming design tailored to low-altitude UAVs, where the antenna array is adapted based on position information to enable robust ISAC integration. %
To address challenges posed by complex propagation environments and constrained resources, \cite{10962171} introduced an MA-aided RIS-aided ISAC architecture. %
This approach jointly optimized antenna placement and RIS reflection coefficients, effectively enhancing spatial diversity and system performance. In \cite{10693833}, the authors examined MA-aided ISAC systems for low-altitude economic applications, providing a holistic analysis encompassing system architecture, performance evaluation, and deployment considerations. %
Finally, \cite{10901248} addressed variations in radar cross section (RCS) among multiple targets and proposed an MA-aided ISAC design optimized for complex multi-target sensing scenarios, significantly enhancing communication and sensing rates in practical environments.


\section{System Model}

\subsection{Transmit Signal Model}

\begin{figure}[htbp]
  \centering
  \includegraphics[scale=0.26]{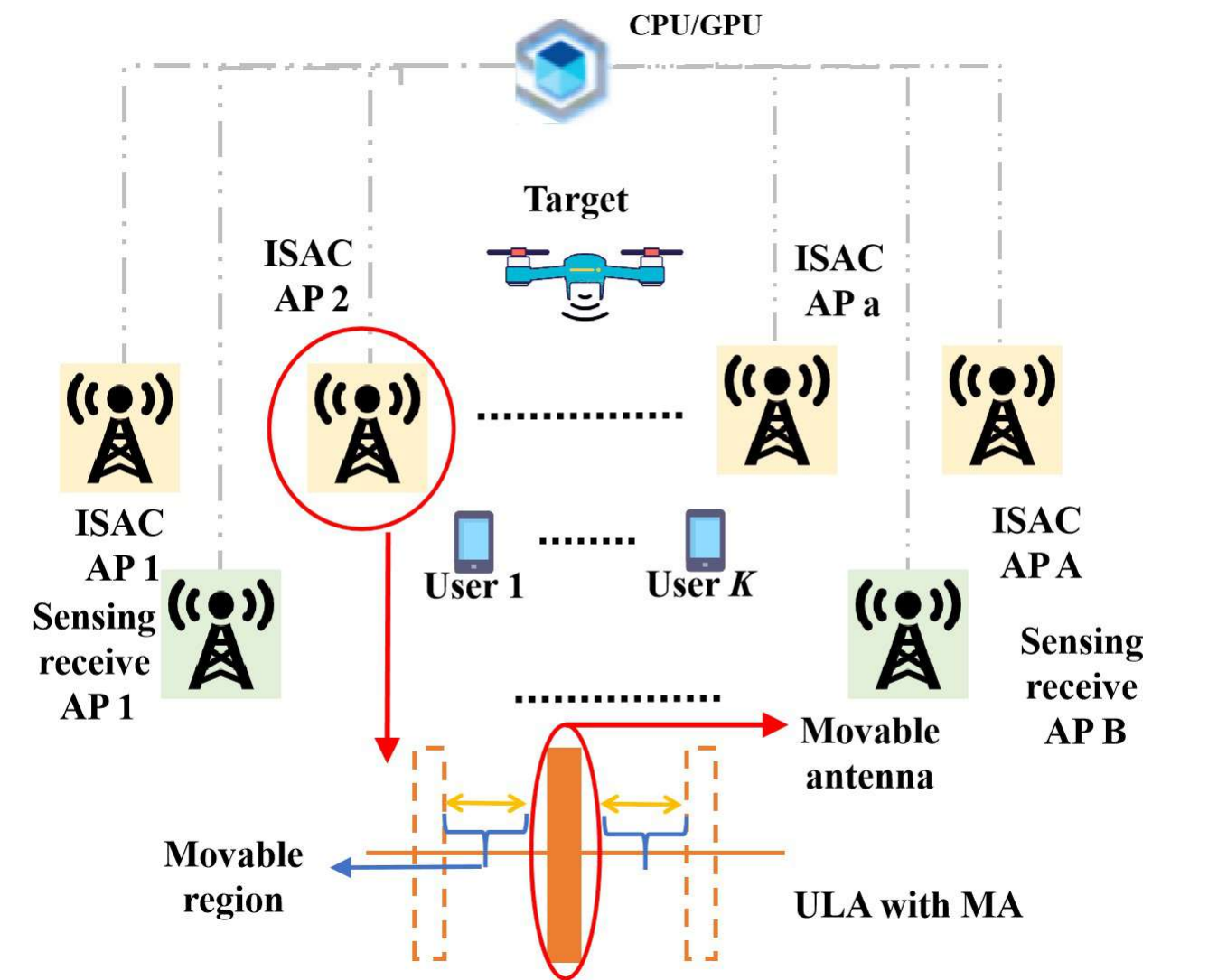}
  \captionsetup{justification=centering}
  \caption{Illustration of the {movable antenna-enhanced} CF-ISAC system.}
\label{FIGURETS1}
\end{figure}

This paper considers an MA-aided CF-ISAC system, as illustrated in Fig.~\ref{FIGURETS1}. In this system, multiple APs collaboratively serve all users while performing target detection.
A CPU is deployed for control and cooperation, with the ISAC transmit APs and sensing receive APs connected to the CPU via optical fiber or wireless backhaul\cite{10516289}. Specifically, the MA-aided CF-ISAC system consists of $A$ ISAC transmit APs, $B$ sensing receive APs, $K$ single-antenna users, and one target. The $a$-th transmit AP and the $b$-th sensing receive AP are equipped with $N_{t}$ MAs and $N_{r}$ MAs, respectively. Furthermore, $S$ denotes the number of time instants. Let 
$\mathcal{A}=\{1,\ldots,A\}$, 
$\mathcal{B}=\{1,\ldots,B\}$, 
$\mathcal{S}=\{0,\ldots,S\}$, and 
$\mathcal{K}=\{1,\ldots,K\}$ denote the index sets of ISAC APs, sensing receive AP, time instants, and users, respectively. %
In addition, the position of the target, the $a$-th ISAC AP, and the $b$-th sensing receive AP are expressed as $\boldsymbol{d}=[d_{x},d_{y}]^{T}$, $\boldsymbol{d}_{a}=[d_{x}^{a},d_{y}^{a}]^{T}$, and $\boldsymbol{d}_{b}=[d_{x}^{b},d_{y}^{b}]^{T}$, respectively. In the proposed MA-aided CF-ISAC system, let
$\boldsymbol{c}_{s}=[c_{s,1},\ldots,c_{s,K}]^{T}\in\mathbb{C}^{K\times 1}$ denote the transmit symbols, where $c_{s,k}$ represents the symbol transmitted to the $k$-th user on the $s$-th time instant. We assume that the transmit symbols satisfy the normalization condition, i.e., 
$\mathbb{E}\{\boldsymbol{c}_{s}\boldsymbol{c}_{s}^{H}\}=\boldsymbol{I}_{K}$,$\forall~s\in\mathcal{S}$. In the downlink, the symbol $c_{s,k}$ is transmitted to the $k$-th user after undergoing beamforming processing with the beamforming vector 
$\boldsymbol{w}_{a,k}\in\mathbb{C}^{N_{t}\times 1}$ at the $a$-th AP. Therefore, the signal $\boldsymbol{x}_{a,s}$ from the $a$-th AP on the $s$-th time instant is denoted as
\begin{align}
\boldsymbol{x}_{a,s}=\sum_{k=1}^{K}\boldsymbol{w}_{a,k}c_{s,k},\label{pro1}
\end{align}

\subsection{TS Errors Model}

Considering TS errors caused by time asynchrony between the ISAC transmit and sensing receive APs, this paper first establishes a propagation delay model. %
Specifically, 
$\tau+\Delta\tau_{a}$ represents the actual transmission time of the ISAC transmitter at the $a$-th AP, and
$\tau+\Delta\tau_{b}$ denotes the actual sensing time of the sensing receiver at the 
$b$-th AP, and $\bar{\tau}_{a,b}$ refers to the TS errors between the ISAC transmitter at the 
$a$-th AP and the sensing receiver at the 
$b$-th AP, which is denoted as  
\begin{align}
\bar{\tau}_{a,b}=\Delta\tau_{a,b}+\nu_{a,b},\label{pro2}
\end{align}
where
$\nu_{a,b}$ represents the bounded measurement noise, with the assumption that 
$\Delta\tau_{a,b}=\Delta\tau_{a}-\Delta\tau_{b}$. According to bounded error model of TS errors in\cite{1599595}, we let $\bar{\tau}_{a,b}\in[\bar{\tau}_{\min},\bar{\tau}_{\max}]$, where $\bar{\tau}_{\min}$ and $\bar{\tau}_{\max}$ represent the minimum TS error and the maximum TS error, respectively.

\subsection{Wideband MA Channel Model}

As illustrated in Fig.~\ref{FIGURETS1}, we consider a downlink multi-user communication scenario in a CF-ISAC system, where multiple APs simultaneously serve multiple users. To characterize the wireless propagation environment between the $k$-th user and the $a$-th AP, we adopt a geometric channel model\cite{10243545}. Specifically, the channel is assumed to consist of $L_{a,k}$ propagation paths. The $l_{a,k}$-th path is associated with a transmit angle of departure (AoD) $\phi_{l_{a,k}}$. In addition, the field response matrix (FRM) of the $a$-th AP is denoted by\cite{10709885} 
\begin{align}
\boldsymbol{g}_{a}(\boldsymbol{p}_{a})=[1,e^{-j2\pi p_{1}^{a}\sin(\phi_{l_{a,k}})},\ldots,e^{-j2\pi p_{N_{t}}^{a}\sin(\phi_{l_{a,k}})}]^{T},\label{pro3}
\end{align}
where $\boldsymbol{p}_{a} = [p_{1}^{a}, \ldots, p_{N_{t}}^{a}]^{T}$ {denotes} the position vector of the transmit MA at the $a$-th AP, in which $p_{t}^{a} \in \mathcal{C}_{a} = [p_{a}^{\min}, p_{a}^{\max}]$ represents the feasible region of movement for the $t$-th antenna element, {for} $1 \leq t \leq N_{t}$. Here, $\mathcal{C}_{a}$ defines the spatial constraints of the MA's movement at the $a$-th AP. Based on the antenna configuration and spatial channel characteristics, the corresponding wideband communication channel model is denoted as
\cite{10032173} 
\begin{align}
\boldsymbol{h}_{a,k}=\sum\nolimits_{l_{a,k}=1}^{L_{a,k}}\alpha_{l_{a,k}}e^{-j2\pi \tau_{l_{a,k}}}\boldsymbol{g}_{a}(\boldsymbol{p}_{a}),\label{pro4}
\end{align}
where {$\alpha_{l_{a,k}}$ denotes the channel gain the $l_{a,k}$-th path, and} $\tau_{l_{a,k}}$ represents the delay associated with the $l_{a,k}$-th channel path component. Following the sensing channel model presented in \cite{7961152}, the corresponding sensing channel is denoted as
\begin{align}
\boldsymbol{H}_{a,b}=\beta_{a,b}\boldsymbol{g}_{b}(\boldsymbol{p}_{b})\bar{\boldsymbol{g}}_{a}^{H}(\boldsymbol{p}_{a}),\label{pro5}
\end{align}
where 
\begin{align}
\boldsymbol{g}_{b}(\boldsymbol{p}_{b})=[1,e^{-j2\pi p_{1}^{b}\sin(\varphi_{b})},\ldots,e^{-j2\pi p_{N_{r}}^{b}\sin(\varphi_{b})}]^{T}\label{pro6}
\end{align}
and
\begin{align}
\bar{\boldsymbol{g}}_{a}(\boldsymbol{p}_{a})=[1,e^{-j2\pi p_{1}^{a}\sin(\varphi_{a})},\ldots,e^{-j2\pi p_{N_{t}}^{a}\sin(\varphi_{a})}]^{T},\label{pro7}
\end{align}
in which $\boldsymbol{p}_{b}=[p_{1}^{b},\ldots,p_{N_{r}}^{b}]^{T}$ denotes the position vector of the receive MA, and each component $p_{r}^{b}$ satisfies $p_{r}^{b}\in\mathcal{C}_{b}=[p_{b}^{\min},p_{b}^{\max}],~1\leq r\leq N_{r}$, where $\mathcal{C}_{b}$ defines the movement region of the receive MA. In addition, in the considered CF-ISAC framework, we model the total propagation delay $\tau_{a,b}$ between the $a$-th AP and the $b$-th AP via the target is given by $\tau_{a,b} = \tau_{a} + \tau_{b}$, where $\tau_{a} = \kappa_{a}/c$ and $\tau_{b} = \kappa_{b}/c$ represent the delays from the $a$-th AP to the target and from the target to the $b$-th AP, respectively. Here, $c$ denotes the speed of light. The distances $\kappa_{a}$ and $\kappa_{b}$ are calculated by
\begin{align}
&\kappa_{a}=\sqrt{(d_{x}^{a}-d_{x})^{2}+(d_{y}^{a}-d_{y})^{2}},\nonumber\\
&\kappa_{b}=\sqrt{(d_{x}^{b}-d_{x})^{2}+(d_{y}^{b}-d_{y})^{2}},
\end{align}
where $(d_{x},d_{y})$ denotes the coordinates of the target, and 
$(d_{x}^{a},d_{y}^{a})$, $(d_{x}^{b},d_{y}^{b})$ are the coordinates of the $a$-th transmit AP and the $b$-th receive AP, respectively. Then,
to enable precise beam alignment and high-resolution sensing, it is essential to characterize the angular parameters of the propagation paths. Accordingly, the AoD and angle of arrival (AoA) associated with the propagation path are computed as
\begin{align}
&\varphi_{a}=\arctan((d_{y}-d_{y}^{a})/(d_{x}-d_{x}^{a}))+\pi\times 1(d_{x}<d_{x}^{a}),\nonumber\\
&\varphi_{b}=\arctan((d_{y}-d_{y}^{b})/(d_{x}-d_{x}^{b}))+\pi\times 1(d_{x}<d_{x}^{b}),
\end{align}
where $1(\cdot)$ is the indicator function, which equals $1$ if the condition is true and $0$ otherwise. These angular metrics provide critical information for directional transmission and sensing in CF-ISAC systems. In this study, the considered MA array is composed of electronically driven equivalent MAs. According to \cite{ning2025movable}, such MAs are capable of adjusting their positions or orientations through high-speed electronic control without requiring mechanical movement. As a result, the movement delay is negligible compared to the overall system latency.

\subsection{Communication System Model}
Based on \cite{6616600} and \cite{4400801}, and assuming proper channel estimation at the communication user, the effect of TS errors on the communication system can be effectively compensated. Therefore, in this work, the impact of TS errors on communication performance is neglected to focus on its influence on the sensing function. The transmitted signal propagates through the wireless channel 
$\boldsymbol{h}_{a,k}$ and is received by user $k$. The transmit signal, after passing through the channel $\boldsymbol{h}_{a,k}$, is received by the user. Thus, receive signal $\boldsymbol{y}_{a,k,s}$ can be expressed as follows
\begin{align}
y_{a,k,s}=\boldsymbol{h}_{a,k}^{H}\boldsymbol{x}_{a,s}=\boldsymbol{h}_{a,k}^{H}\sum\nolimits_{k=1}^{K}\boldsymbol{w}_{a,k}c_{s,k}.\label{pro8}%
\end{align}
There are $A$ APs simultaneously serving 
$K$ users, and the receive signal at user $k$ is the sum of the signals transmitted by 
$A$ APs. Let $y_{k,s}$ denote the receive signal at user $k$ on subcarrier $s$. Considering the additive white Gaussian noise (AWGN) at the receiver, the receive signal is given by 
\begin{align}
&y_{k,s}=\sum\nolimits_{a=1}^{A}\boldsymbol{y}_{a,k,s}+\boldsymbol{z}_{k,s}=\sum\nolimits_{a=1}^{A}\boldsymbol{h}_{a,k}^{H}\sum\nolimits_{k=1}^{K}\boldsymbol{w}_{a,k}c_{s,k}\nonumber\\
&+z_{k,s}
=\underbrace{\sum\nolimits_{a=1}^{A}\boldsymbol{h}_{a,k}^{H}\boldsymbol{w}_{a,k}c_{s,k}}_{\text{{desired} signal}}\nonumber\\
&+\underbrace{\sum\nolimits_{a=1}^{A}\sum\nolimits_{i=1,i\neq k}^{K}\boldsymbol{h}_{a,k}^{H}\boldsymbol{w}_{a,i}c_{s,i}}_{\text{interference signal}}+z_{k,s},\label{pro9}
\end{align}
where 
$\boldsymbol{z}_{k,s}$ represents the AWGN with {mean zero} and covariance $\sigma_{k,s}^{2}\boldsymbol{I}$. The first term in {(\ref{pro9})} represents the {desired} signal at user 
$k$, while the second term accounts for the {co-channel interference signal intended for} other users. Since there are 
$S$ available subcarriers in total, the receive signal at user $k$ is denoted as 
$\{y_{k,s}\}_{s=1}^{S}$.
Then, the SINR for the transmitted symbol $c_{s,k}$ at the user $k$ on {subcarrier} $s$ can {easily be} evaluated, {which then yields the weighted} sum-rate (WSR) $R_{sum}$ of all $K$ users {as}
\begin{align}
&R_{sum}=\sum\nolimits_{k=1}^{K}\sum\nolimits_{s=1}^{S}\eta_{k}R_{k,s}=\sum\nolimits_{k=1}^{K}\sum\nolimits_{s=1}^{S}\nonumber\\
&\eta_{k}\log_{2}\left(1+\frac{|\sum_{a=1}^{A}\boldsymbol{h}_{a,k}^{H}\boldsymbol{w}_{a,k}|^{2}}{\left(\sum_{j=1,j\neq k}^{K}|\sum_{a=1}^{A}\boldsymbol{h}_{a,k}^{H}\boldsymbol{w}_{a,j}|^{2}+\sigma_{k,s}^{2}\right)}\right),\label{pro10}
\end{align}
where $\eta_{k}\in\mathbb{R}^{+}$ represents the weight of {user} $k$ and $R_{k,s}$ denotes the rate of user $k$ {on} subcarrier $s$.

\subsection{Sensing System Model}

In this section, we focus on estimating the target position $\boldsymbol{d}$ and employ the CRLB to assess the accuracy of target position estimation. The presence of TS errors, which cannot be compensated at the sensing receiver, introduces ambiguity in TS errors measurements, resulting in errors in target position estimation. Therefore, it is crucial to investigate the position estimation performance under TS errors, which motivates the derivation of the CRLB incorporating TS errors to establish a precise performance bound. Specifically, we assume that each sensing {receive} AP collects signals from $\bar{S}$ time slots during the signal duration $S$. The receive time-domain signal is first converted to the frequency-domain signal. Let $\bar{c}_{\bar{s},i}$ denote the baseband frequency-domain symbol transmitted by AP $a$ on frequency-domain sample $\bar{s}$. Then, the receive frequency-domain signal at user $k$ on the $\bar{s}$-th frequency-domain sample is represented as
\begin{align}
\boldsymbol{y}_{b,\bar{s}}=\sum\nolimits_{a=1}^{A}\boldsymbol{H}_{a,b}\sum\nolimits_{i=1}^{K}\boldsymbol{w}_{a,i}\bar{c}_{\bar{s},i}e^{-jf_{\bar{s}}\bar{\tau}_{a,b}}+\bar{\boldsymbol{z}}_{b,\bar{s}},\label{pro11}
\end{align}
where $\bar{\boldsymbol{z}}_{b,\bar{s}}$ is the AWGN. Then, the signal from all subcarriers is given by
\begin{align}
\boldsymbol{y}_{b}=\sum\nolimits_{a=1}^{A}\boldsymbol{\Xi}_{a,b}\boldsymbol{x}_{a}+\bar{\boldsymbol{z}}_{b},\label{pro12}
\end{align}
where 
\begin{small}
\begin{align}
&{\boldsymbol{\Xi}_{a,b}=\mathrm{Blkdiag}[\boldsymbol{H}_{a,b},\ldots,\boldsymbol{H}_{a,b}]},\nonumber\\
&{\boldsymbol{x}_{a}=\mathrm{vec}(\boldsymbol{X}_{b})(\mathrm{diag}(\boldsymbol{\vartheta}_{a,b})\otimes\boldsymbol{1})=\left[\left(\sum\nolimits_{k=1}^{K}\boldsymbol{w}_{a,k}\bar{c}_{1,k}e^{-jf_{1}\bar{\tau}_{a,b}}\right)^{T}\right.}\nonumber\\
&
{,\ldots,\left.\left(\sum\nolimits_{k=1}^{K}\boldsymbol{w}_{a,k}\bar{c}_{\bar{S},k}e^{-jf_{\bar{S}}\bar{\tau}_{a,b}}\right)^{T}\right]^{T}},\nonumber\\
&{\boldsymbol{X}_{b}=\left[\left(\sum\nolimits_{k=1}^{K}\boldsymbol{w}_{a,k}\bar{c}_{1,k}\right),\ldots,\left(\sum\nolimits_{k=1}^{K}\boldsymbol{w}_{a,k}\bar{c}_{\bar{S},k}\right)\right]^{T}},\nonumber\\
&\boldsymbol{\vartheta}_{a,b}=[e^{-jf_{1}\bar{\tau}_{a,b}},\ldots,e^{-jf_{\bar{S}}\bar{\tau}_{a,b}}]^{T},\nonumber\\
&{\boldsymbol{z}_{b}=[(\boldsymbol{z}_{b,1})^{T},\ldots,(\boldsymbol{z}_{b,\bar{S}})^{T}]^{T}}.\\
&\boldsymbol{y}_{b}=[(\boldsymbol{y}_{b,1})^{T},\ldots,(\boldsymbol{y}_{b,\bar{S}})^{T}]^{T}
\end{align}
\end{small}%
Based on (\ref{pro12}), we can rewrite the signal from all subcarriers as
\begin{align}
\boldsymbol{y}_{b}=\sum\nolimits_{a=1}^{A}\boldsymbol{\Xi}_{a,b}\mathrm{vec}(\boldsymbol{X}_{b})(\mathrm{diag}(\boldsymbol{\vartheta}_{a,b})\otimes\boldsymbol{1})+\boldsymbol{z}_{b}.\label{pro13}
\end{align}
Then, we can obtain the CRLB for target position $\boldsymbol{d}$, {which yields} the CRLB of the position parameter {as}
\begin{align}
\mathrm{\textbf{CRLB}}_{b}(\boldsymbol{d})=(\boldsymbol{F}_{\boldsymbol{d}\boldsymbol{d}}^{b})^{-1},\label{pro14}
\end{align}
where 
$\boldsymbol{F}_{\boldsymbol{d}\boldsymbol{d}}^{b}$ represents the {Fisher information matrix (FIM)} and the detailed expression is given in \textbf{Appendix~A}. Finally, all mathematical symbols are defined in \textbf{Table}~\ref{tab:notation}.

\begin{table}[h!]
\centering
\begin{tabular}{|l|l|}
\hline
\textbf{Symbol} & \textbf{Description} \\ \hline
$\mathcal{A}/\mathcal{B}/\mathcal{K}/\mathcal{S}$ & Set of transmit APs/receive APs/subcarriers/users \\ \hline
$A/B/K/S$ & Number of transmit APs/receive APs/subcarriers/users \\ \hline
$N_{t}/N_{r}$ & Number of transmit MAs/receive MAs \\ \hline
$\boldsymbol{d}/\boldsymbol{d}_{a}/\boldsymbol{d}_{b}$ & Position of target/the $a$th transmit AP/the $b$th receive AP \\ \hline
$\boldsymbol{c}_{s}$ & Transmit symbol \\ \hline
$\boldsymbol{w}_{a,k}$ & Transmit beamforming vector \\ \hline
$\bar{\tau}_{a,b}$ & TS errors between the $a$th transmit AP and $b$th\\
~&sensing receive AP \\ \hline
$\boldsymbol{p}_{a}/\boldsymbol{p}_{b}$ & Position of transmit MAs/receive MAs\\ \hline
$\mathcal{C}_{a}/\mathcal{C}_{b}$ & Movable region of transmit MAs/receive\\
&MAs \\ \hline
$\boldsymbol{H}_{a,b}$ & Sensing channel\\ \hline
$\boldsymbol{h}_{a,k,s}$ & Communication channel \\ \hline
$L_{a,k}$ & Number of channel paths \\ \hline
$\varphi_{a}$/$\varphi_{b}$ & AoA/AOD of sensing channel\\\hline
$D_{0}$ & Minimum distance of MAs\\ \hline
\end{tabular}
\caption{Summary of {Notations}.}
\label{tab:notation}
\end{table}

\section{Problem Formulation}
Each diagonal element of the CRLB matrix represents the minimum variance of the corresponding parameter estimated by an unbiased estimator. Therefore, the trace of the CRLB matrix is utilized to characterize the accuracy of target estimation\cite{10477314}. Our objective is to maximize the CRLB of target estimation in the presence of TS errors by designing the positions of the MAs at the APs and coordinating the transmit beamforming, while satisfying the communication rate and transmit power constraints. Notably, to achieve a robust joint design for the system, we focus on maximizing the worst-case CRLB by optimizing the TS errors  $\bar{\tau}_{a,b}$, ensuring that the proposed design remains resilient to uncertainties arising from varying TS errors. Specifically, the problem of maximizing sensing accuracy, subject to the communication rate constraint, can be formulated as 
\begin{subequations}
\begin{align}
\min_{\boldsymbol{w}_{a,k},\boldsymbol{p}_{a},\boldsymbol{p}_{b}}\max_{\bar{\tau}_{a,b}}&~\sum\nolimits_{b=1}^{B}\mathrm{\textbf{CRLB}}_{b}(\boldsymbol{d}),\label{pro15a}\\
\mbox{s.t.}~
&\bar{\tau}_{min}\leq\bar{\tau}_{a,b}\leq\bar{\tau}_{max},&\label{pro15b}\\
&R_{k,s}\geq\gamma_{k,s},&\label{pro15c}\\
&p_{t}^{a}\in\mathcal{C}_{a}, 1\leq t\leq N_{t},&\label{pro15d}\\
&p_{r}^{b}\in\mathcal{C}_{b}, 1\leq r\leq N_{r},&\label{pro15e}\\
&|p_{t}^{a}-p_{t-1}^{a}|\geq D_{0},&\label{pro15f}\\
&|p_{r}^{b}-p_{r-1}^{b}|\geq D_{0},&\label{pro15g}\\
&\|\boldsymbol{w}_{a,k}\|_{2}^{2}\leq P_{max},&\label{pro15h}
\end{align}\label{pro15}%
\end{subequations}
where (\ref{pro15b}) denotes the constraint on the TS errors, (\ref{pro15c}) represents the communication rate constraint. The constraints 
(\ref{pro15d}) and 
(\ref{pro15e}) define the feasible movement regions of the transmitting and receiving MAs, respectively. 
(\ref{pro15f}) and 
(\ref{pro15g}) define the minimum movable distance $D_{0}$ between the transmitted and received MAs. Finally, 
(\ref{pro15h}) denotes the transmit power constraint imposed on the system.
This section addresses the robust joint optimization problem (\ref{pro15}). The problem (\ref{pro15}) can be decomposed into two subproblems
\begin{subequations}
\begin{align}
\max_{\bar{\tau}_{a,b}}&~\sum\nolimits_{b=1}^{B}\mathrm{\textbf{CRLB}_{b}}(\boldsymbol{d}),\label{pro16a}\\
\mbox{s.t.}~
&(\ref{pro15b})&\label{pro16b}
\end{align}\label{pro16}
\end{subequations}
and
\begin{subequations}
\begin{align}
\min_{\boldsymbol{w}_{a,k},\boldsymbol{p}_{a},\boldsymbol{p}_{b}}&~\sum\nolimits_{b=1}^{B}\mathrm{\textbf{CRLB}_{b}}(\boldsymbol{d}),\label{pro17a}\\
\mbox{s.t.}~
&(\ref{pro15c}),(\ref{pro15d}),(\ref{pro15e}),(\ref{pro15f}),(\ref{pro15g}),(\ref{pro15h}).&\label{pro17b}
\end{align}\label{pro17}%
\end{subequations}
However, due to the coupling of the optimization variables $\bar{\tau}_{a,b}$, $\boldsymbol{w}_{a,k}$, $\boldsymbol{p}_{a}$ and $\boldsymbol{p}_{b}$ in the objective function, problems (\ref{pro16}) and (\ref{pro17}) are difficult to solve. To overcome this challenge, we propose an efficient algorithm based on the alternating optimization (AO) framework. Specifically, for the worst-case CRLB problem, we apply an MO algorithm to solve the problem (\ref{pro16}). Then, with the optimization errors of TS, we employ MA-MetaRL to solve the problem (\ref{pro17}).

\section{Manifold Optimization-based Algorithm to Problem (\ref{pro16})}
According to the definition of manifolds provided in \cite{9472958,10797657}, the TS error vector 
$\boldsymbol{\vartheta}=[\boldsymbol{\vartheta}_{1,1}^{T},\ldots,\boldsymbol{\vartheta}_{A,1}^{T},\ldots,$ $\boldsymbol{\vartheta}_{1,B}^{T},\ldots,\boldsymbol{\vartheta}_{A,B}^{T}]$ is regarded as an embedded submanifold of the Euclidean space 
$\mathbb{C}^{ABS\times 1}$, where the TS error vector is represented by 
\begin{align}
\mathcal{O}=\{\boldsymbol{\vartheta}\in\mathbb{C}^{ABS\times 1}:|\vartheta_{j}|=1,j=1,2,\ldots,ABS\},\label{pro18}
\end{align}
in which the vector $\vartheta_{j}$ is the 
$j$-th element of $\boldsymbol{\vartheta}$. In general, a manifold 
$\mathcal{O}$ is not as naturally suited for optimization as Euclidean or vector spaces. To overcome this limitation, a Riemannian manifold is introduced, characterized by an inner product that varies smoothly across the tangent spaces at each point\cite{absil2010optimization}. Similar to how the derivative of a complex-valued function provides a local linear approximation, the tangent space 
$\mathcal{T}_{\boldsymbol{\vartheta}}\mathcal{O}$  at point 
$\boldsymbol{\vartheta}$ serves as a local vector space approximation of the manifold 
$\mathcal{O}$. Moreover, the presence of an inner product enables the definition of various geometric concepts on the manifold. By further treating 
$\mathbb{C}$ as a space 
$\mathbb{R}^{2}$ equipped with a canonical inner product, the Euclidean metric 
$\mathbb{C}$ on the complex plane can be defined accordingly as
\begin{align}
<\vartheta_{1},\vartheta_{2}>=\mathcal{R}[\vartheta_{1},\vartheta^{*}_{2}].\label{pro19}
\end{align}
Based on the definition of the inner product, the tangent vector of $\mathcal{O}$ can be defined accordingly. For a vector $\boldsymbol{\vartheta}$, if the inner product between each element in $\boldsymbol{\vartheta}$ and its corresponding element in $\boldsymbol{\vartheta}$ is $0$, then $\boldsymbol{\vartheta}$ is said to be orthogonal to 
$\boldsymbol{\psi}$, i.e.,
\begin{align}
<\psi_{j},\vartheta_{j}>=0,\forall~j,\label{pro20}
\end{align}
where the element 
$\boldsymbol{\psi}_{i}$ is the 
$j$-th element of the set 
$\boldsymbol{\psi}$. By interpreting each complex-valued element of a vector as a vector in 
$\mathbb{R}^{2}$, 
$\boldsymbol{\psi}$ can be regarded as the tangent vector of 
$\mathcal{O}$ at point 
$\boldsymbol{\vartheta}$, provided that the following condition is satisfied
\begin{align}
\mathrm{Re}[\boldsymbol{\psi}\odot\boldsymbol{\vartheta}^{*}]=\boldsymbol{0}.\label{pro21}
\end{align}
The tangent space of the manifold 
$\mathcal{O}$ at point 
$\boldsymbol{\vartheta}$, denoted by 
$\mathcal{T}_{\boldsymbol{\vartheta}}\mathcal{O}$, is the set of all tangent vectors of 
$\mathcal{O}$ at point 
$\boldsymbol{\vartheta}$. Accordingly, the tangent space of 
$\boldsymbol{\vartheta}$ is expressed as
\begin{align}
\mathcal{T}_{\boldsymbol{\vartheta}}\mathcal{O}=\{\boldsymbol{\psi}\in\mathbb{C}^{S\times 1}:\mathrm{Re}[\boldsymbol{\psi}\odot\boldsymbol{\vartheta}^{*}]=\boldsymbol{0}\}.\label{pro22}
\end{align}
Owing to the local homeomorphism between neighborhoods on a manifold and Euclidean space\cite{absil2010optimization}, optimization algorithms originally developed for Euclidean spaces can be locally extended to Riemannian manifolds. By leveraging the vector space structure provided by the tangent space, conventional line search methods can be effectively applied within this framework. Based on this principle, this paper proposes a Riemannian MO-based gradient algorithm for TS error vector $\boldsymbol{\vartheta}$.
Based on (\ref{pro15}), we define the cost function as 
\begin{align}
L(\boldsymbol{\vartheta})=\sum\nolimits_{b=1}^{B}\mathrm{Tr}\{\textbf{CRLB}_{b}(\boldsymbol{d})\}.\label{pro23}
\end{align}
To minimize the CRLB at each AP, it is necessary to determine the direction of steepest descent of the cost function within the current tangent space. Under the Riemannian optimization framework, the Euclidean gradient is first computed as 
\begin{align}
\frac{\partial L(\boldsymbol{\vartheta})}{\partial \boldsymbol{\vartheta}}=\frac{\bar{\boldsymbol{C}}D-C\bar{\boldsymbol{D}}}{D^{2}},\label{pro24}
\end{align}
where $C$, $D$, $\bar{\boldsymbol{C}}$ and $\bar{\boldsymbol{D}}$ are given in \textbf{Appendix~B}. Then projected orthogonally onto the tangent space to obtain the Riemannian gradient, and the projection is given by
\begin{align}
P_{\boldsymbol{\vartheta}}\left(\frac{\partial L(\boldsymbol{\vartheta})}{\partial \boldsymbol{\vartheta}}\right)=\frac{\partial L(\boldsymbol{\vartheta})}{\partial \boldsymbol{\vartheta}}-\mathrm{R}\left\{\frac{\partial L(\boldsymbol{\vartheta})}{\partial \boldsymbol{\vartheta}}\odot\boldsymbol{\vartheta}\right\}\odot\boldsymbol{\vartheta}.\label{pro25}
\end{align}
In Euclidean space, directions of movement are naturally represented by vectors. However, when dealing with curved spaces, such as spheres or more general manifolds, a global linear structure no longer exists, and direction cannot be globally defined using vectors in the same way. To address this limitation, a tangent space is introduced at each point on the manifold, serving as a local linear approximation that captures all possible directions in which one can move from that point.
Accordingly, the movement of a point $\boldsymbol{\vartheta}$ is interpreted as taking place within its associated tangent space $\mathcal{T}_{\boldsymbol{\vartheta}}\mathcal{O}$, which provides a rigorous mathematical framework for defining and analyzing directional motion on manifolds\cite{absil2010optimization}. In addition, on a manifold, this idea is generalized by the concept of a retraction, which enables movement along a tangent vector while ensuring that 
$\boldsymbol{\vartheta}$ remains on the manifold 
$\mathcal{O}$. Given a point 
$\boldsymbol{\vartheta}$ on 
$\mathcal{O}$, a step size 
$\alpha$, and a search direction 
$\bar{\boldsymbol{d}}$, the new point is computed as follows:
\begin{align}
\mathrm{Ret}_{\boldsymbol{\vartheta}}(\alpha\boldsymbol{d})=\left[\frac{\vartheta_{1}+\alpha \bar{d}_{1}}{|\vartheta_{1}+\alpha \bar{d}_{1}|},\ldots,\frac{\vartheta_{ABS}+\alpha \bar{d}_{S}}{|\vartheta_{ABS}+\alpha \bar{d}_{S}|}\right]^{T}.\label{pro26}
\end{align}
By iteratively computing the gradient and updating the point on the manifold, we develop a conjugate gradient algorithm for TS error vector $\boldsymbol{\vartheta}$. After obtaining the TS error vector, the equality constraint in (\ref{pro18}) is transformed into 
$\sum_{a=1}^{A}\sum_{b=1}^{B}|\angle\boldsymbol{\vartheta}_{a,b}[s]+f_{s}\bar{\tau}_{a,b}|^{2}$, and the corresponding optimization problem for computing $\bar{\tau}_{a,b}$ is formulated as 
\begin{subequations}
\begin{align}
\min_{\bar{\tau}_{a,b}}\sum\nolimits_{a=1}^{A}\sum\nolimits_{b=1}^{B}&~\sum\nolimits_{s=1}^{S}|\angle\boldsymbol{\vartheta}_{a,b}[s]+f_{s}\bar{\tau}_{a,b}|^{2},\label{pro27a}\\
\mbox{s.t.}~
&\bar{\tau}_{min}\leq\bar{\tau}_{a,b}\leq\bar{\tau}_{max}.&\label{pro27b}
\end{align}\label{pro27}%
\end{subequations}
Since problem (\ref{pro27}) is convex, the optimization problem can be solved {by using the CVX tool}\cite{gb08}. The algorithm is summarized in \textbf{Algorithm~\ref{algo1}}.
\begin{algorithm}%
\caption{Manifold-based AO algorithm for (\ref{pro16})} \label{algo1}
\hspace*{0.02in}{\bf Input:} $\bar{\tau}_{a,b}^{(0)}$.\\
\hspace*{0.02in}{\bf Repeat:}~$n=n+1$.\\
Computing Armijo backtracking line search step size $\alpha_{t}$ \cite{absil2010optimization}.\\
Computing TS error vector based on (\ref{pro26})\\
Updating Reimannian gradient $\boldsymbol{u}^{(n+1)}=P_{\boldsymbol{\vartheta}^{(n+1)}}\left(\frac{\partial L(\boldsymbol{\vartheta})}{\partial \boldsymbol{\vartheta}}|_{\boldsymbol{\vartheta}=\boldsymbol{\vartheta}^{(n+1)}}\right)$.\\
Updating Polak-Ribiere parameter based on (\ref{pro25}).\\
Computing search direction $\bar{\boldsymbol{d}}^{(n+1)}=\boldsymbol{u}^{(n+1)}+\beta^{(n+1)}P_{\boldsymbol{\vartheta}^{(n+1)}}(\bar{\boldsymbol{d}}^{(n)})$.\\
\hspace*{0.02in}{\bf Until:} Obtain the local optimal solution $\boldsymbol{\vartheta}^{*}$.\\ 
\hspace*{0.02in}{\bf Repeat:}~$\bar{n}=\bar{n}+1$.\\
Solving the problem in (\ref{pro27}) by using CVX.\\
\hspace*{0.02in}{\bf Until:} Obtain the local optimal solution $\bar{\tau}_{a,b}^{*}$.\\
\hspace*{0.02in}{\bf Until:} The above two algorithms are iteratively executed until convergence is achieved.\\
\end{algorithm}

%




\section{Meta Reinforcement Learnin algorithm to Problem (\ref{pro17})}\label{IV}

Model‑free reinforcement learning (RL) offers a principled way to tackle sequential decision problems without an explicit system model.  
Because the mobility of the mobile agents (MAs) makes the CF‑ISAC environment highly non‑stationary, we cast Problem~(\ref{pro17}) as a Markov decision process (MDP) and solve it with a meta-reinforcement learning (MetaRL) framework built on an enhanced TD3 learner.

\subsection{Problem Reformulation as an MDP}\label{sec:IV-A}
We cast Problem~(\ref{pro17}) as a \emph{continuous} MDP
\((\mathcal{S},\mathcal{A},P,R,\gamma)\).  
In the following, we describe each component in detail.

\textbf{Action space \(\mathcal{A}\).}
At the beginning of slot \(j\) the controller jointly determines  
(i) the transmit beamforming vector of every AP  
and (ii) the positions of the MAs.  
Let \(\boldsymbol{w}_{a,k}\in\mathbb{C}^{N_t}\) be the complex beamforming vector of AP~\(a\).  
We convert it into a real vector by stacking its real and imaginary parts:
\(
  \check{\boldsymbol{w}}_{a,k}
  \triangleq
  [\Re\{\boldsymbol{w}_{a,k}\}^{T},
   \Im\{\boldsymbol{w}_{a,k}\}^{T}]^{T}\in\mathbb{R}^{2N_t}.
\)
Denote the MA positions by \(\boldsymbol{p}_a\in\mathbb{R}^{N_{t}},\boldsymbol{p}_b\in\mathbb{R}^{N_{r}}\).
The \emph{continuous} action vector is therefore
\begin{equation}
   \boldsymbol{a}_j^T
   =
   \bigl[
     \check{\boldsymbol{w}}_{a,1}^{T}\!\cdots
     \check{\boldsymbol{w}}_{a,K}^{T},
     \boldsymbol{p}_a^{T},
     \boldsymbol{p}_b^{T}
   \bigr]
   \;\in\;
   \mathcal{A}=\mathbb{R}^{\,(2K+1)N_t+N_{r}}.
   \label{eq:action_vec}
\end{equation}
Because \(\mathcal{A}\) is high‑dimensional and real‑valued, we later adopt the
TD3 algorithm as base learner.

\textbf{State space \(\mathcal{V}\).}
Each state aggregates instantaneous environment measurements as well as the
most recent action–reward pair, providing the agent with temporal context:
\begin{equation}
   \boldsymbol{s}_j^T
   =
   \bigl[
     \underbrace{\boldsymbol{e}_j^{T}}_{\text{environment}},
     \underbrace{\boldsymbol{a}_{j-1}^{T}}_{\text{prev.\ action}},
     \underbrace{r_{j-1}}_{\text{prev.\ reward}}
   \bigr]
   \in\mathcal{V}.
   \label{eq:state_vec}
\end{equation}
The environment feature vector
\(
  \boldsymbol{e}_j
  =
  \bigl[
     R_{1,1},\ldots,R_{K,V},
     P_{\max},D_0
  \bigr]^{T}
\)
contains the instantaneous signal‑to‑noise ratios \(R_{k,v}\) between
AP~\(a\) and user, the maximum transmit power \(P_{\max}\), and the guard distance \(D_0\).

\textbf{Transition kernel \(P\).}
The unknown probability
\(P(\boldsymbol{s}_{j+1}\!\mid\!\boldsymbol{s}_j,\boldsymbol{a}_j)\) is induced by
radio‑channel variation, user mobility, and MA motion.  
We therefore adopt a \emph{model‑free} learning strategy that infers
optimal behaviour solely from sampled transitions.

\textbf{Reward function \(R\).}
The objective is to minimise the CRLB
while satisfying all system constraints.
Let \(\operatorname{tr}(\mathbf{CRLB}_b)\) be the trace of the CRLB matrix at AP~\(b\)
and let \(C_1\) denote the number of constraints.
Introduce the indicator
\(
  \mathbf{1}_{\text{vio},i}(\boldsymbol{s}_j,\boldsymbol{a}_j)=
  1\) if constraint \(i\) is violated and \(0\) otherwise.
We define the \emph{scalar} reward as
\begin{equation}
   r(\boldsymbol{s}_j,\boldsymbol{a}_j)
   =
   -\sum_{b=1}^{B}\operatorname{tr}(\mathbf{CRLB}_b)
   \;-\;
   \beta\sum_{i=1}^{C_1}\mathbf{1}_{\text{vio},i},
   \quad
   \beta>0.
   \label{eq:reward_def}
\end{equation}
Thus higher reward corresponds to a lower aggregate CRLB and zero
constraint violation.

\textbf{Policy \(\pi_\mu\).}
We employ a \emph{stochastic} policy
\(
   \pi_\mu(\boldsymbol{a}_j\!\mid\!\boldsymbol{s}_j)
\)
parameterised by a neural network~\(\mu\).
During execution the agent samples
\(
  \boldsymbol{a}_j\sim\pi_\mu(\cdot\mid\boldsymbol{s}_j)
\),
while during training TD3 gradually adjusts~\(\mu\)
to maximise the expected discounted return
\(
  \mathbb{E}\!\bigl[\sum_{k=0}^{\infty}\gamma^{k}\,
  r(\boldsymbol{s}_{j+k},\boldsymbol{a}_{j+k})\bigr]
\)
with discount factor \(\gamma\in(0,1]\).

With the continuous spaces \(\mathcal{S}\) and \(\mathcal{A}\) fully specified,
Problem~(\ref{pro17}) is now an MDP amenable to deep RL.
The next subsection introduces the MA‑MetaRL framework that
combines an enhanced TD3 learner with meta‑learning to
handle fast environmental changes caused by MA position.

\subsection{MA--MetaRL Framework}\label{sec:IV-B}
\begin{figure}[htbp]
  \centering
  \includegraphics[width=0.45\textwidth, height=0.31\textwidth]{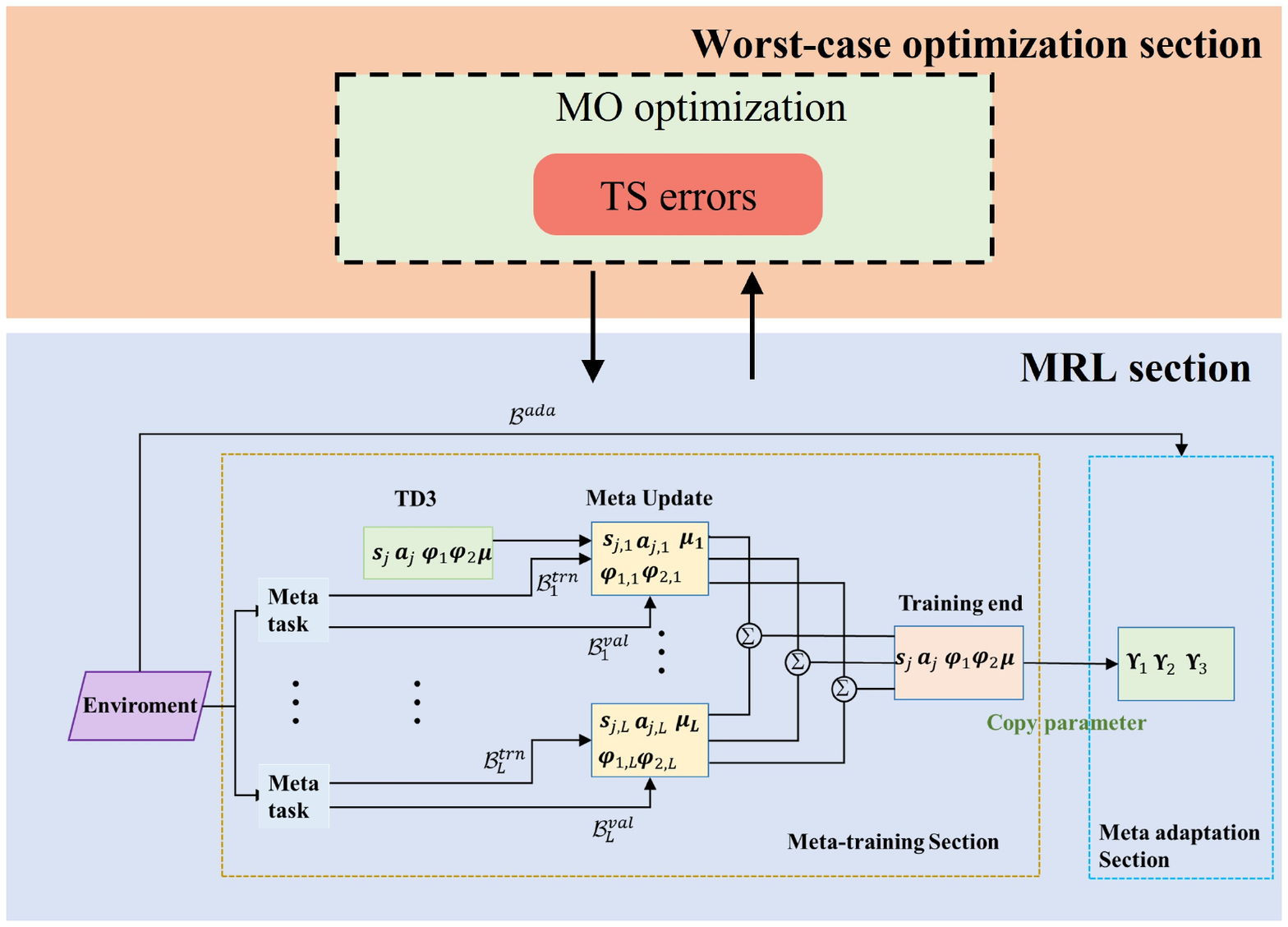}
  \captionsetup{justification=centering}
  \caption{Flow chart of proposed MA-MetaRL algorithm.}
\label{FIGUREFlow}
\end{figure}

Every random placement of users and sensing targets defines a distinct MDP, so a policy trained on one layout generalises poorly to others.  We address this with the MA-MetaRL algorithm which is described in \textbf{Algorithm 2}, and the flow chart of the algorithm is given in \textbf{Figure.~\ref{FIGUREFlow}}.  It couples the enhanced TD3 learner with model‑agnostic meta‑reinforcement learning.  Offline, \(L\) layouts are sampled; each constitutes a task \(\mathcal{T}_{\ell}\) with its own replay buffer \(\mathcal{B}_{\ell}\).  For each task the agent executes a few inner‑loop TD3 updates on \(\mathcal{B}_{\ell}^{\text{trn}}\), evaluates the adapted parameters on \(\mathcal{B}_{\ell}^{\text{val}}\), and aggregates the resulting gradients to refine a single set of meta‑initial weights \((\boldsymbol{\mu},\boldsymbol{\varphi}_{1},\boldsymbol{\varphi}_{2})\) that maximise post‑adaptation return averaged across all tasks.  At run time, facing an unseen layout \(\mathcal{T}_{\text{new}}\), the agent performs the same brief TD3 update on its newly collected buffer \(\mathcal{B}^{\text{adp}}\), thereby obtaining a layout‑specialised policy after only a handful of interactions.

\begin{algorithm}[!ht]
\caption{Proposed MA-MetaRL algorithm for (\ref{pro25})} \label{algo2}
\hspace*{0.02in}{\bf Initialize:}
$\boldsymbol{\varphi}_{1}$, $\boldsymbol{\varphi}_{2}$, $\boldsymbol{\mu}$,$J$,$L$,$N$ episodes,$\bar{\boldsymbol{\varphi}}_{1}\longleftarrow\boldsymbol{\varphi}_{1}$, $\bar{\boldsymbol{\varphi}}_{2}\longleftarrow\boldsymbol{\varphi}_{2}$, $\tilde{\boldsymbol{\varphi}}_{1}^{l}\longleftarrow\boldsymbol{\varphi}_{1}$, $\tilde{\boldsymbol{\varphi}}_{2}^{l}\longleftarrow\boldsymbol{\varphi}$,$\tilde{\boldsymbol{\mu}}^{l}\longleftarrow\boldsymbol{\mu}$, $\mathcal{B}_{l}$,$\mathcal{B}_{adp}$.\\
\hspace*{0.02in}{\bf Repeat:}~$n=n+1$.\\
Select state $\boldsymbol{s}_{j}$ and action $\boldsymbol{a}_{j}$.\\
Compute action based on the environment, and obtain the reward $r(\boldsymbol{s}_{j},\boldsymbol{a}_{j})$.\\
Store $(\boldsymbol{s}_{j},\boldsymbol{a}_{j},r(\boldsymbol{s}_{j},\boldsymbol{a}_{j}),\boldsymbol{s}_{j+1})$ in $\mathcal{B}$\\
Update the network parameters $\boldsymbol{\varphi}_{1}$, $\boldsymbol{\varphi}_{2}$ and $\boldsymbol{\mu}$ based on (\ref{pro33}) and (\ref{pro35}).\\
Update the network parameters $\bar{\boldsymbol{\varphi}}_{1}$, $\bar{\boldsymbol{\varphi}}_{2}$ and $\bar{\boldsymbol{\mu}}$ based (\ref{pro38})\\
\hspace*{0.02in}{\bf Repeat:}~$n=n+1$.\\
\hspace*{0.02in}{\bf Meta-training phase:} Executive (\ref{pro37}) and (\ref{pro38}) to achieve the Meta-training.\\
\hspace*{0.02in}{\bf Meta-adaptation phase:} Executive (\ref{pro38}) to achieve the Meta-adaptation.\\
\end{algorithm}

\subsubsection{Base Learner: Enhanced TD3}\label{sec:IV-C}

The TD3 algorithm is a model-free, off-policy RL method built upon the Actor–Critic framework. Its primary objective is to interact with the environment iteratively to optimize the policy such that the expected long-term cumulative reward is maximized\cite{fujimoto2018addressing}. In this context, we employ the state–action value function to quantify the expected return of taking a specific action in a given state and is formally defined as follows:
\begin{align}
&q_{\mu}(\boldsymbol{s}_{j},\boldsymbol{a}_{j})=\mathbb{E}_{P(\boldsymbol{s}_{j+1}|\boldsymbol{s}_{j},\boldsymbol{a}_{j})}\left[\sum\nolimits_{j=0}^{\infty}\gamma^{j}r(\boldsymbol{s}_{j},\mu(\boldsymbol{s}_{j})|\boldsymbol{s}_{j}=\boldsymbol{s}_{0}\right.\nonumber\\
&\left.,\boldsymbol{a}_{j}=\mu(\boldsymbol{s}_{0}))\right],\label{pro30}
\end{align}
where $\gamma\in(0,1]$ denotes the discount factor. The optimal policy is given by
\begin{align}
\mu^{*}(\boldsymbol{s}_{j})=\arg\max_{\mu(\boldsymbol{s}_{j})\in\mathcal{A}} q_{\mu}(\boldsymbol{s}_{j},\mu(\boldsymbol{s}_{j})).\label{pro31}
\end{align}
The TD3 algorithm is composed of six DNNs, including an actor network (AN) with parameters $\boldsymbol{\mu}$, two critic networks (CN1 and CN2) with parameters $\boldsymbol{\varphi}_{1}$ and $\boldsymbol{\varphi}_{2}$, respectively, a target actor network (TAN) with parameters $\bar{\boldsymbol{\mu}}$, and two target critic networks (TCN1 and TCN2) with parameters $\bar{\boldsymbol{\varphi}}_{1}$ and $\bar{\boldsymbol{\varphi}}_{2}$. As an extension of the deep deterministic policy gradient (DDPG) framework, TD3 incorporates several structural enhancements aimed at mitigating the overestimation of the state–action value function, thereby reducing the risk of converging to suboptimal policies. These improvements are elaborated in detail in the training procedure section.

During training, batches of data $\mathcal{K}(\boldsymbol{s}_{j,\bar{k}},\boldsymbol{a}_{j,\bar{k}},r(\boldsymbol{s}_{j,\bar{k}},\boldsymbol{a}_{j,\bar{k}}),\boldsymbol{s}_{j+1,\bar{k}})$ are randomly sampled from the experience replay buffer $\mathcal{B}$ to update the networks, where $\bar{k}\in\{1,\ldots,B\}$, and $B$ denotes the batch size. The input $\boldsymbol{s}_{j,\bar{k}}$ is fed into the AN to produce an action $\boldsymbol{a}_{j,\bar{k}}$. CN1 and CN2 receive inputs $\mathcal{K}_{\boldsymbol{\mu}}(\boldsymbol{s}_{j,\bar{k}},\boldsymbol{a}_{j,\bar{k}};\boldsymbol{\varphi}_{1})$ and $\mathcal{K}_{\boldsymbol{\mu}}(\boldsymbol{s}_{j,\bar{k}},\boldsymbol{a}_{j,\bar{k}};\boldsymbol{\varphi}_{2})$, respectively, to compute value estimates $\boldsymbol{s}_{j+1,\bar{k}}$ and $\boldsymbol{a}_{j+1,\bar{k}}$. Simultaneously, TAN takes the input $\boldsymbol{s}_{j,\bar{k}}$ from the replay buffer and generates an action $\boldsymbol{a}_{j,\bar{k}}$. To improve robustness against erroneous value estimates, the output of TAN is perturbed by adding noise—an essential modification in TD3—to smooth the target state-action value function. The final smoothed target action is given by
\begin{align}
\tilde{\boldsymbol{a}}_{j+1,\bar{k}}=\mathrm{clip}(\bar{\boldsymbol{\mu}}(\boldsymbol{s}_{j+1,\bar{k}})+\mathrm{clip}(\varepsilon^{\prime},-v,v),a_{min},a_{max}),\label{pro32}
\end{align}
where $\varepsilon^{\prime}\sim\mathcal{N}(0,\sigma)$ is the sample noise, and $v$ denote the maximum value of sample noise.  TCN1 and TCN2 receive the state $\boldsymbol{s}_{j+1,\bar{k}}$ and action $\tilde{\boldsymbol{a}}_{j+1,\bar{k}}$ as inputs, producing outputs $\mathcal{K}_{\bar{\boldsymbol{\mu}}}(\boldsymbol{s}_{j+1,\bar{k}},\bar{\boldsymbol{a}}_{j+1,\bar{k}}:\bar{\boldsymbol{\varphi}}_{1})$ and $\mathcal{K}_{\bar{\boldsymbol{\mu}}}(\boldsymbol{s}_{j+1,\bar{k}},\bar{\boldsymbol{a}}_{j+1,\bar{k}}:\bar{\boldsymbol{\varphi}}_{2})$, respectively. To mitigate the overestimation bias commonly encountered in value function approximation, the minimum of the two target Q-value estimates is adopted as the learning target for updating the critic networks CN1 and CN2 with parameters $\boldsymbol{\varphi}_1$ and $\boldsymbol{\varphi}_2$, thereby enhancing the stability and reliability of policy learning. The loss function used to update the parameters of CN1 and CN2 is defined as follows:
\begin{align}
&\mathcal{G}(\boldsymbol{\varphi}_{1})=\frac{1}{|B|}\sum\nolimits_{\bar{k}=1}^{B}(\mathcal{K}_{\boldsymbol{\mu}}(\boldsymbol{s}_{j,\bar{k}},\boldsymbol{a}_{j,\bar{k}};\boldsymbol{\varphi}_{1})-y(r_{j,k},\boldsymbol{s}_{j+1,,\bar{k}}))^{2},\nonumber\\
&\mathcal{G}(\boldsymbol{\varphi}_{2})=\frac{1}{|B|}\sum\nolimits_{\bar{k}=1}^{B}(\mathcal{K}_{\boldsymbol{\mu}}(\boldsymbol{s}_{j,\bar{k}},\boldsymbol{a}_{j,\bar{k}};\boldsymbol{\varphi}_{2})-y(r_{j,\bar{k}},\boldsymbol{s}_{j+1,\bar{k}}))^{2},\label{pro33}
\end{align}
where
\begin{align}
&y(r_{j,\bar{k}},\boldsymbol{s}_{j+1,\bar{k}})=r_{j,\bar{k}}+\gamma\min_{\boldsymbol{\varphi}_{1},\boldsymbol{\varphi}_{2}}\{\mathcal{K}_{\boldsymbol{\mu}}(\boldsymbol{s}_{j,\bar{k}},\boldsymbol{a}_{j,\bar{k}};\boldsymbol{\varphi}_{1}),\nonumber\\
&\mathcal{K}_{\boldsymbol{\mu}}(\boldsymbol{s}_{j,\bar{k}},\boldsymbol{a}_{j,\bar{k}};\boldsymbol{\varphi}_{2})\}.
\end{align}
To update the parameters $\boldsymbol{\varphi}_{1}$ and $\boldsymbol{\varphi}_{2}$ of CN1 and CN2, the loss function in (\ref{pro33}) is optimized using a gradient descent algorithm, denoted as 
\begin{align}
&\boldsymbol{\varphi}_{1}=\boldsymbol{\varphi}_{1}-\theta_{1}\nabla_{\boldsymbol{\varphi}_{1}}\mathcal{G}(\boldsymbol{\varphi}_{1}), \boldsymbol{\varphi}_{2}=\boldsymbol{\varphi}_{2}-\theta_{2}\nabla_{\boldsymbol{\varphi}_{2}}\mathcal{G}(\boldsymbol{\varphi}_{2}),\label{pro34}
\end{align}
in which $0<\theta_{1}<1$ and $0<\theta_{2}<1$ represent the learning rate. The third modification in the TD3 algorithm involves updating the parameters of AN and TAN less frequently than those of the other networks. When computing the loss function for AN, only $\mathcal{K}_{\bar{\boldsymbol{\mu}}}(\boldsymbol{s}_{j,\bar{k}},\bar{\boldsymbol{a}}_{j,\bar{k}}:\bar{\boldsymbol{\varphi}}_{1})$ is used, as defined below
\begin{align}
\mathcal{G}(\boldsymbol{\mu})=\frac{1}{|B|}\sum\nolimits_{k=1}^{B}\mathcal{K}_{\bar{\boldsymbol{\mu}}}(\boldsymbol{s}_{j,k},\bar{\boldsymbol{a}}_{j,k}:\bar{\boldsymbol{\varphi}}_{1}).\label{pro35}
\end{align}
Gradient descent can also be used to update AN parameters
\begin{align}
\boldsymbol{\mu}=\boldsymbol{\mu}-\theta_{3}\nabla_{\mu}\mathcal{L}(\boldsymbol{\mu}).
\end{align}
The TAN, TCN1, and TCN2 are refreshed based on the AN, CN1, and CN2
\begin{align}
&\bar{\boldsymbol{\varphi}}_{1}=\tau_{\boldsymbol{\varphi}}\boldsymbol{\varphi}_{1}+(1-\tau_{\alpha})\bar{\boldsymbol{\varphi}}_{1},\bar{\boldsymbol{\varphi}}_{2}=\tau_{\boldsymbol{\varphi}}\boldsymbol{\varphi}_{2}+(1-\tau_{\alpha})\bar{\boldsymbol{\varphi}}_{2},\nonumber\\
&\bar{\boldsymbol{\mu}}=\tau_{\boldsymbol{\mu}}\boldsymbol{\mu}+(1-\tau_{\boldsymbol{\mu}})\bar{\boldsymbol{\mu}},\label{pro36}
\end{align}
where $\tau_{\alpha}$ and $\tau_{\boldsymbol{\mu}}$ are the decaying rates for the AN and
CNs, respectively. 

\subsubsection{Meta‑Training and Meta‑Adaptation}\label{sec:IV-D}

The proposed MA-MetaRL algorithm consists of two main phases: meta-training and meta-adaptation. During the meta-training phase, the model is trained by randomly sampling a batch set, denoted as $\mathcal{B}_{l}^{trn}$, from $\mathcal{B}_{l}$. The parameters of the TD3 Actor and CNs for each task are updated using the stochastic gradient descent (SGD) algorithm\cite{10032173}. Meanwhile, another batch set, $\mathcal{B}_{l}^{val}$, is used to update the global model parameters for all tasks $L$.

In the meta-adaptation phase, a new agent is introduced to evaluate the performance of MA-MetaRL. For TD3, the agent consists of an AN, two CNs, a TAN, and two TCNs. During training, a new task is defined, and the user is randomly placed. Unlike the meta-training phase, the model parameters in the meta-adaptation phase are initialized using parameters obtained from the meta-training phase, rather than random initialization. Additionally, the meta-adaptation phase involves storing state transitions, including the current state, next state, action, and reward, in the experience memory $\mathcal{B}_{adp}$.

During the meta-training phase, each task $l$ begins by resetting the environment to obtain the initial state $\boldsymbol{s}_{1}^{l}$. At each time step $j$, the agent selects an action $\boldsymbol{a}_{j}^{l}$ based on the current policy and executes it, receiving a reward $r(\boldsymbol{s}_{j}^{l},\boldsymbol{a}_{j}^{l})$ and transitioning to the next state $\boldsymbol{s}_{j+1}^{l}$. The resulting experience tuple $(\boldsymbol{s}_{j}^{l},\boldsymbol{s}_{j+1}^{l},\boldsymbol{a}_{j}^{l},r(\boldsymbol{s}_{j}^{l},\boldsymbol{a}_{j}^{l}))$, $\mathcal{B}_{l}$, and $\mathcal{B}_{l}$ is stored in the replay buffer $R$. Once the number of stored transitions reaches the predefined batch size, a mini-batch $\mathcal{B}_{l}^{trn}$ is randomly sampled from the experience pool $\mathcal{B}_{l}^{trn}$. This mini-batch $\mathcal{B}_{l}$ is then used to update the AN and CN parameters of TD3 for each task $l$, denoted as $\tilde{\boldsymbol{w}}_{a,s,b}$, $\boldsymbol{x}_{a}$, and $\boldsymbol{y}_{b}$, respectively. The loss function used to train the DNNs is defined as
\begin{align}
&\tilde{\boldsymbol{\mu}}_{l}=\arg\min_{\mu} \mathcal{G}_{l}(\boldsymbol{\mu},\mathcal{B}_{l}^{trn}),\tilde{\boldsymbol{\varphi}}_{1}^{l}=\arg\min_{\varphi_{1}} \mathcal{G}_{l}(\boldsymbol{\varphi}_{1},\mathcal{B}_{l}^{trn}),\nonumber\\
&\tilde{\boldsymbol{\varphi}}_{2}^{l}=\arg\min_{\varphi_{2}} \mathcal{G}_{l}(\boldsymbol{\varphi}_{2},\mathcal{B}_{l}^{trn}).\label{pro37}
\end{align}
After completing the steps for all tasks 
$l\in\mathcal{L}$, the evaluation phase is performed. Specifically, a batch $\mathcal{B}_{l}$ is sampled from 
$\mathcal{B}_{l}^{val}$. The TD3 is updated using the gradients of the corresponding loss functions $\sum_{l}\mathcal{G}_{l}(\boldsymbol{\mu},\mathcal{B}_{l}^{trn})$, $\sum_{l}\mathcal{G}_{l}(\boldsymbol{\varphi}_{1},\mathcal{B}_{l}^{trn})$, and 
$\sum_{l}\mathcal{G}_{l}(\boldsymbol{\varphi}_{2},\mathcal{B}_{l}^{trn})$. Finally, the optimization problems for 
$\boldsymbol{\mu}$, $\boldsymbol{\varphi}_{1}$, and $\boldsymbol{\varphi}_{2}$ can be expressed as
\begin{align}
&\boldsymbol{\mu}=\arg\min_{\mu} \sum_{l}\mathcal{G}_{l}(\hat{\boldsymbol{\mu}}_{l},\mathcal{B}_{l}^{val}),\boldsymbol{\varphi}_{1}=\arg\min_{\varphi_{1}}\sum_{l} \mathcal{G}_{l}(\hat{\boldsymbol{\varphi}}_{1,l},\mathcal{B}_{l}^{val}),\nonumber\\
&\boldsymbol{\varphi}_{2}=\arg\min_{\varphi_{2}}\sum_{l} \mathcal{G}_{l}(\hat{\boldsymbol{\varphi}}_{2,l},\mathcal{B}_{l}^{val}).\label{pro38}
\end{align}
In the meta-training phase, a learning model has been trained for all meta-tasks. During the meta-adaptation phase, the parameters of the MA-MetaRL algorithm are used to train a new learning model that specifically adapts to the new task. Similar to the meta-training phase, the meta-adaptation phase utilizes an AN and two CNs, with parameters corresponding to $\boldsymbol{\Upsilon}_{1}$, $\boldsymbol{\Upsilon}_{2}$, and $\boldsymbol{\Upsilon}_{3}$ in the TD3 algorithm. In the initial step, the network parameters are replaced with the corresponding parameters from the training phase model, the update expressions are expressed as
\begin{align}
&\boldsymbol{\Upsilon}_{1}=\boldsymbol{\Upsilon}_{1}-\gamma_{1}\frac{\partial \mathcal{G}_{l}(\boldsymbol{\Upsilon}_{1},\mathcal{B}^{ada})}{\partial \boldsymbol{\Upsilon}_{1}}, \boldsymbol{\Upsilon}_{2}=\boldsymbol{\Upsilon}_{2}-\gamma_{2}\frac{\partial \mathcal{G}_{l}(\boldsymbol{\Upsilon}_{2},\mathcal{B}^{ada})}{\partial \boldsymbol{\Upsilon}_{2}},\nonumber\\ 
&\boldsymbol{\Upsilon}_{3}=\boldsymbol{\Upsilon}_{3}-\gamma_{3}\frac{\partial \mathcal{G}_{l}(\boldsymbol{\Upsilon}_{3},\mathcal{B}^{ada})}{\partial \boldsymbol{\Upsilon}_{3}},\label{pro40}
\end{align}
in which $\gamma_{1}$, $\gamma_{2}$, and $\gamma_{3}$ represent the learning rate of the AN and CN.

\section{Numerical Results}\label{V}
\begin{figure}[htbp]
  \centering
  \includegraphics[scale=0.3]{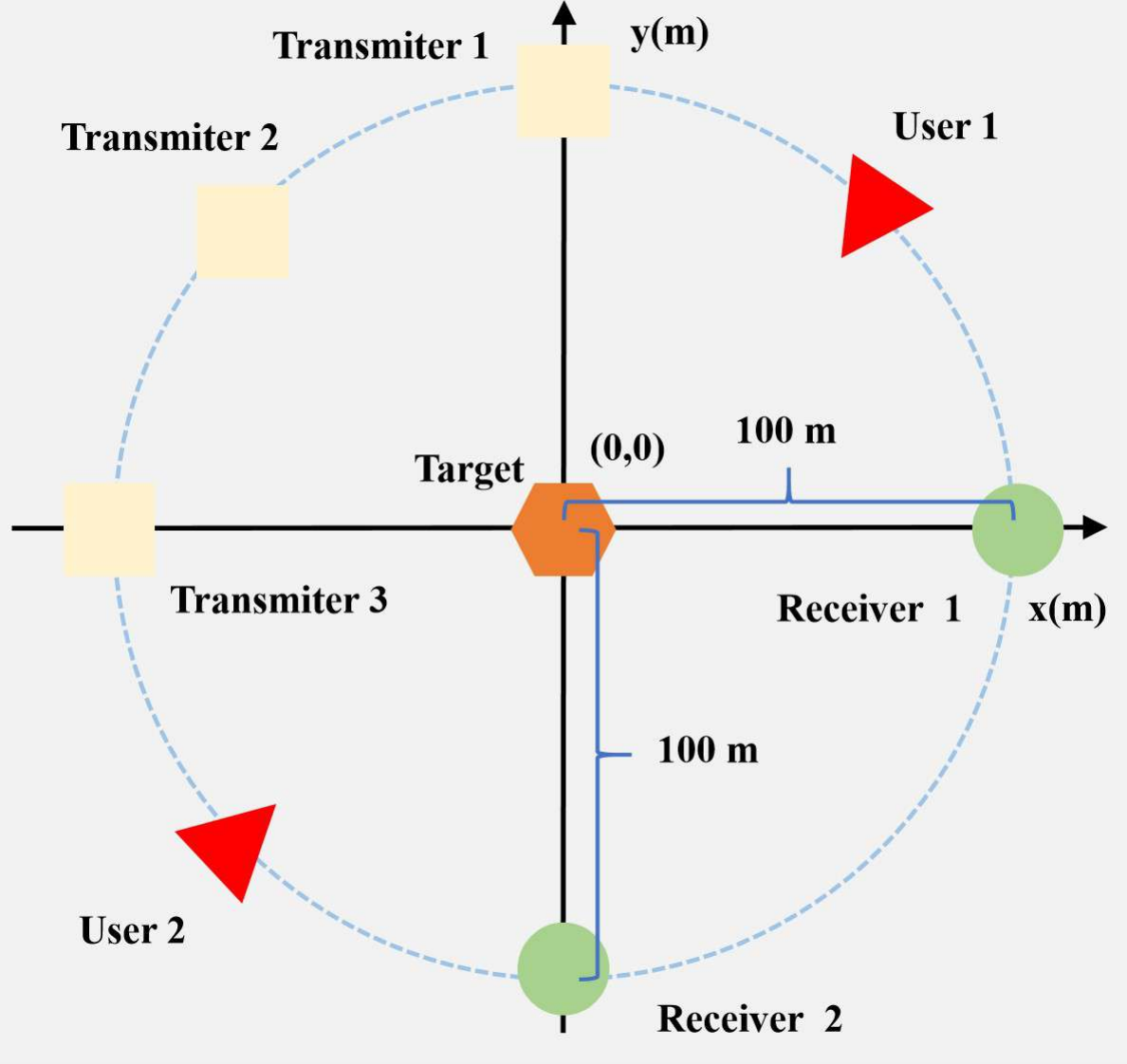}
  \captionsetup{justification=centering}
  \caption{Simulation setup of the MA-aided CF-ISAC system.}
\label{FIGURETS2}
\end{figure}
In the simulation setup, the sensing target is located at the origin of the coordinate system. The ISAC APs, sensing receiver APs, and users are uniformly distributed along a circular ring with a radius of $100$ meters, as illustrated in Fig.~\ref{FIGURETS2}. The system parameters are configured as follows: the number of ISAC APs $A = 3$, sensing APs $B = 2$, and users $U = 2$. Each AP is equipped with {$N_{t} = 16$} transmit MAs and $M = 4$ receive MAs. The channel path loss is modeled by $PL(d) = PL_{0}(d/d_{0})^{-\Omega}$, where $PL_{0} = -30$ dB denotes the reference path loss {at} $d_{0} = 1$ m, $d$ represents the link distance, and $\Omega$ is the path loss exponent. The number of channel paths is set to $L_{b,u} = 3$, and the noise power level is fixed at $-80$ dBm. The minimum spacing between adjacent MAs is $D_{0} = \lambda/2$. The transmit MA {movable range} is defined {at} $x_{t,a}^{\min} = -2\lambda$ and $x_{t,a}^{\max} = 2\lambda$, while the receive MA {movable range} is set to $y_{r,b}^{\min} = -2\lambda$ and $y_{r,b}^{\max} = 2\lambda$. The path loss exponents are set to $2.8$ for the AP-user links and $2.2$ for the AP-target links. The RCS of the target is $\alpha = 3$, and the sensing accuracy is specified as $\gamma_{b} = 0.05$.
To evaluate the performance of
the proposed MA-MetaRL robust algorithm, 
we consider the six benchmark schemes for performance
comparison, which are:
\begin{itemize}
    \item \textbf{Baseline 1:} {Proposed MA-MetaRL algorithm without considering TS errors (ideal case).}
    \item \textbf{Baseline 2:} Proposed DRL algorithm without TS errors roust algorithm\cite{10480601}.
    \item \textbf{Baseline 3:} Proposed soft actor-critic (SAC) algorithm without TS errors {robust} algorithm\cite{10480601}.
    \item \textbf{Successive Convex Approximation (SCA):} SCA-based algorithm for beamforming and MA positions optimization\cite{10709885}.
    \item \textbf{FPA algorithm:} SCA-based algorithm for beamforming optimization with FPA structure\cite{10709885}.
\end{itemize}

\begin{figure}[htbp]
\centering
\begin{minipage}[t]{0.48\textwidth}
\centering
\includegraphics[width=0.9\textwidth, height=0.65\textwidth]{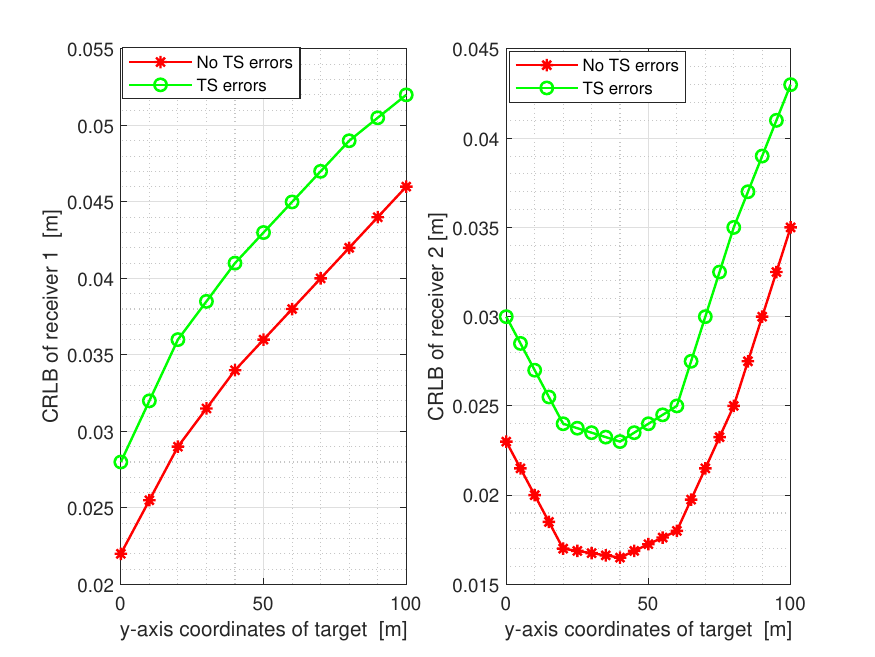}
\put(-150,-5){\small\textbf{(a)}}
\put(-60,-5){\small\textbf{(b)}}
\caption{(a) Impact of TS errors on overall sensing CRLB of sensing receiver 1. (b) Impact of TS errors on overall sensing CRLB of sensing receiver 2.}
\label{FIGURETS3}
\end{minipage}
\end{figure}

As illustrated in Fig.\ref{FIGURETS3}(a) and Fig.\ref{FIGURETS3}(b), the CRLB in the presence of TS errors varies depending on the target's position relative to the receivers. In Fig.\ref{FIGURETS3}(a), the CRLB of receiver $1$ increases with the target’s distance, and the performance gap between the cases with and without TS errors increases. This is {because as} the target moves farther from Receiver 1, the increased signal propagation distance {results in} reduced SNR {further making} the system more sensitive to TS errors. These factors amplify the uncertainty caused by TS errors. Moreover, \ref{FIGURETS3}(b) shows that as the target moves away from the origin, the CRLB at Receiver 2 decreases due to improved SNR resulting from a reduced transmission distance. However, as the target moves beyond a certain distance, both CRLB curves (with and without TS errors) start to increase again, due to the dominant effect of path loss and the resulting degradation in signal quality with increasing distance.

\begin{figure}[htbp]
  \centering
  \includegraphics[scale=0.45]{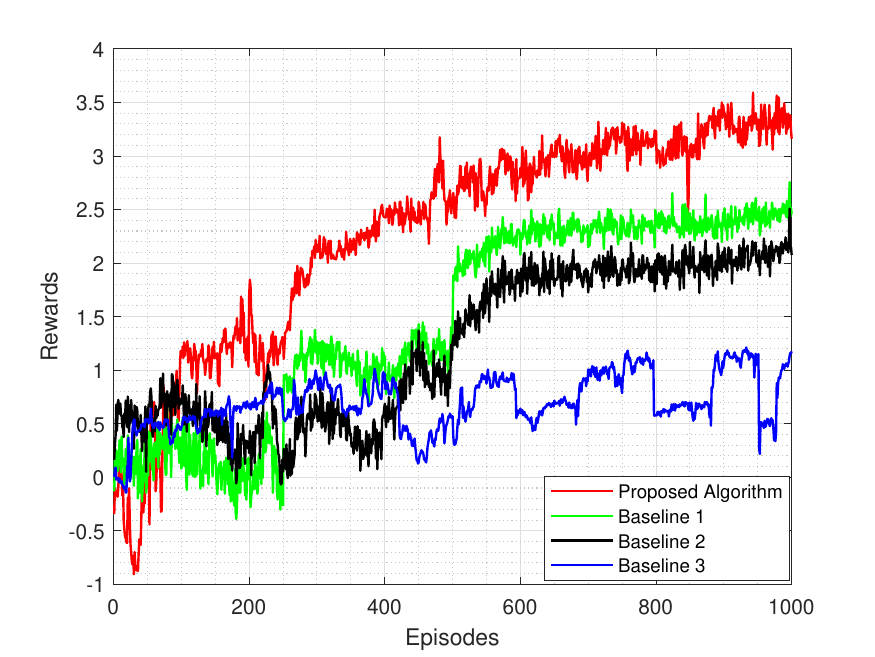}
  \captionsetup{justification=centering}
  \caption{Reward convergence analysis of the proposed MA-MetaRL and baselines.}
\label{FIGURETS4}
\end{figure}


\begin{figure}[htbp]
  \centering
  \includegraphics[scale=0.45]{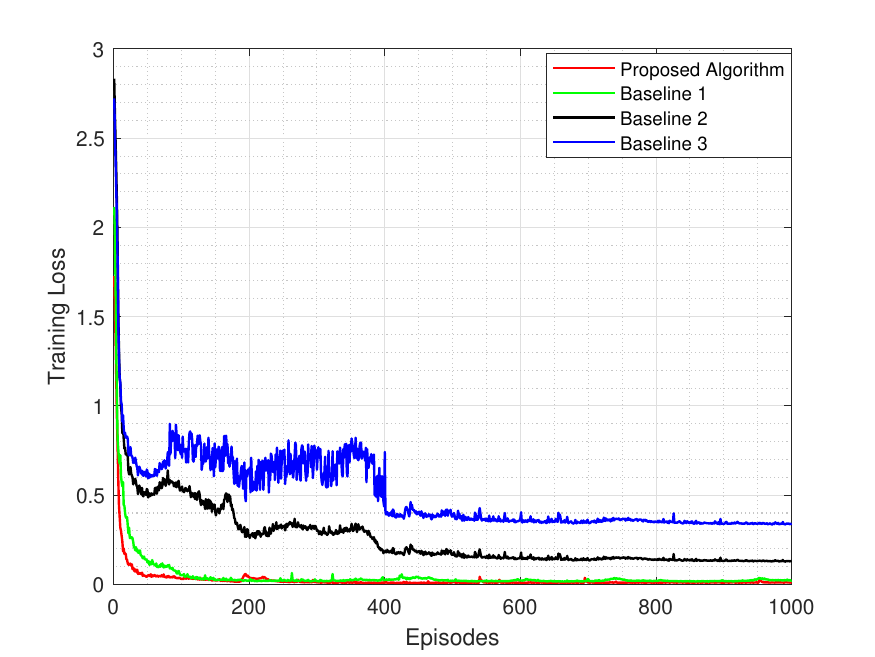}
  \captionsetup{justification=centering}
  \caption{Training loss convergence analysis of the proposed MA-MetaRL and baselines.}
\label{FIGURETS5}
\end{figure}

As shown in Fig.~\ref{FIGURETS4}, as the number of training episodes increases, all four algorithms, including our proposed method and the three baselines, exhibit gradual convergence in rewards, but our proposed algorithm demonstrates significantly faster convergence speed. The key reason lies in the fact that the three baseline algorithms do not incorporate robust processing for TS errors, resulting in a complex and enlarged action space due to uncertainties caused by TS errors. This increases the difficulty in policy learning and causes unstable gradient updates, leading to slow and fluctuating convergence. In contrast, our proposed approach integrates a TS errors robust mechanism into the MA-MetaRL framework, effectively modeling timing uncertainty and reducing the complexity of the decision space. By explicitly accounting for TS-induced distortions during both state representation and policy optimization, the algorithm avoids learning suboptimal actions caused by TS errors.

As shown in Fig.~\ref{FIGURETS5}, as the number of training episodes increases, the training loss for all algorithms gradually converges, reflecting improved policy learning over time. However, our proposed algorithm consistently achieves significantly lower training loss compared to the three baseline methods. This is primarily due to the incorporation of a robust treatment for TS errors and designed MA-MetaRL framework that effectively constrains the action space, enabling more efficient and stable policy updates. In contrast, the three baseline algorithms do not consider robustness to TS errors, resulting in an excessively large and unstructured action space that degrades learning efficiency. This leads to unstable policy exploration and convergence behavior, leading to much higher and fluctuating training loss throughout the episodes. Our approach’s integration of TS errors modeling and robust policy optimization ensures better generalization and faster convergence, highlighting its advantage in both learning efficiency.

\begin{figure}[htbp]
\centering
\begin{minipage}[t]{0.48\textwidth}
\centering
\includegraphics[width=0.9\textwidth, height=0.65\textwidth]{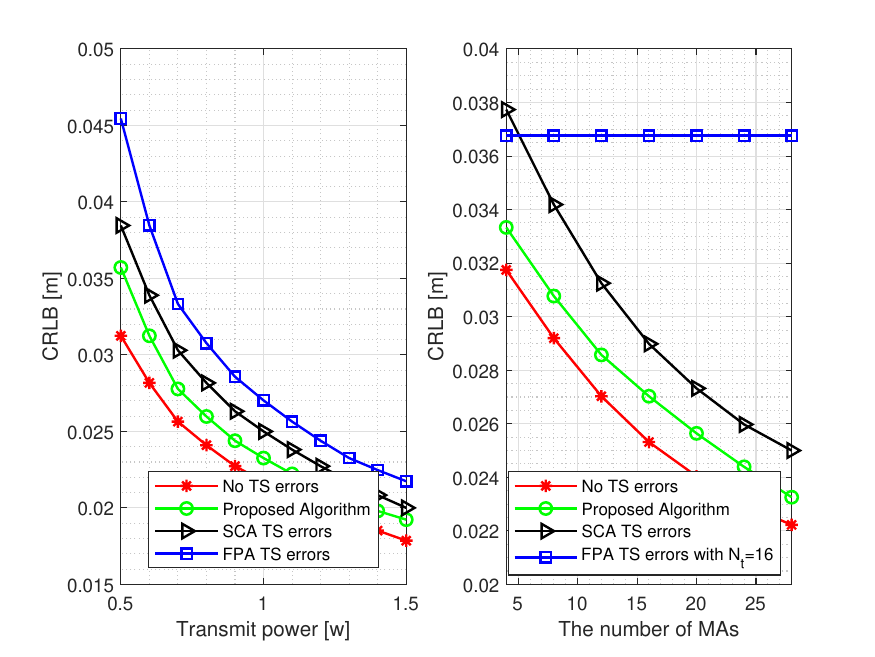}
\put(-150,-5){\small\textbf{(a)}}
\put(-60,-5){\small\textbf{(b)}}
\caption{(a) CRLB versus the transmit power. (b) CRLB versus the number of MAs.}
\label{FIGURETS6}
\end{minipage}
\end{figure}

Fig.~\ref{FIGURETS6}(a) and Fig.~\ref{FIGURETS6}(b) illustrate the sensing CRLB under different system parameters. In Fig.~\ref{FIGURETS6}(a), as the transmit power increases, all algorithms, including the \textbf{Baseline 1}, the proposed algorithm, the \textbf{SCA-based algorithm}, and the \textbf{FPA-based} method with TS errors, show a decreasing trend in CRLB. This is because a higher transmit power improves the overall signal strength, thereby improving the effective SINR. Among these, the proposed MA-MetaRL algorithm consistently outperforms SCA-based and FPA methods due to its ability to jointly optimize antenna positions and beamforming under TS errors, thus better mitigating the adverse impact of TS errors. In Fig.~\ref{FIGURETS6}(b), as the number of MAs increases, the CRLB of the no-TS error scheme, the proposed MA-MetaRL algorithm and the SCA-based method all decrease, while the CRLB of the FPA-based method remains unchanged. This is because MAs provide additional spatial DoF that can be exploited to enhance channel quality and spatial diversity. In contrast, the FPA lacks the ability to adapt to the channel environment. The superior performance of the proposed MA-MetaRL approach over the conventional convex-optimization-based SCA method.

\begin{figure}[htbp]
\centering
\begin{minipage}[t]{0.48\textwidth}
\centering
\includegraphics[width=0.9\textwidth, height=0.65\textwidth]{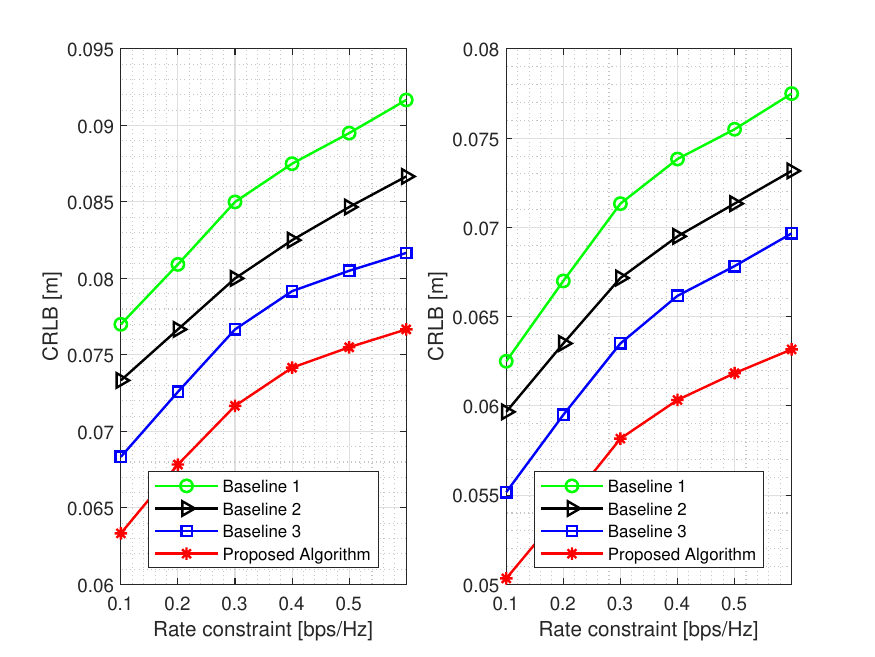}
\put(-150,-5){\small\textbf{(a)}}
\put(-60,-5){\small\textbf{(b)}}
\caption{(a) CRLB versus the transmit rate constraint with TS error interval $[0.4ns, 0.6ns]$. (b) CRLB versus the transmit rate constraint with TS error interval $[0.5ns, 0.8ns]$.}
\label{FIGURETS7}
\end{minipage}
\end{figure}

The simulation results are illustrated in Fig.~\ref{FIGURETS7}. As shown in Fig.~\ref{FIGURETS7}(a), under a TS error interval $[0.4ns, 0.6ns]$, the CRLB degrades as $\gamma_{k,s}$ gradually increases. The reason lies in the fact that $\gamma_{k,s}$ provides greater flexibility in balancing sensing and communication performance. This allows the system to trade a modest transmit rate for a substantial gain in CRLB. In addition, Fig.~\ref{FIGURETS7}(b) illustrates the scenario where the TS error interval increases to $[0.5 ns, 0.8ns]$. Although relaxing $\gamma_{k,s}$ still yields performance gains across all algorithms, the CRLB declines slowly. The slow degradation is attributed to the larger TS error, which introduces significant temporal misalignment and mismatches in signal reception. These impairments substantially weaken the spatial gain benefits of beamforming and multi-antenna cooperation, leading to a reduction in the effective SINR. In particular, in both Fig.~\ref{FIGURETS7}(a) and Fig.~\ref{FIGURETS7}(b), the proposed MA-MetaRL algorithm consistently outperforms conventional SCAmethods.  This is because the SCA method is prone to being trapped in local optima. In contrast, MA-MetaRL explores the solution space through long-term interactions and policy optimization, enabling it to escape local minima and discover better solutions. Compared to DRL, the improvement in CRLB of MA-MetaRL stems from its ability to mitigate overestimation bias through the use of twin Q-networks. Moreover, MA-MetaRL further enhances robustness by learning shared policy structures across tasks, enabling stable performance across multiple non-stationary optimization problems.

\section{Conclusion}\label{VI}
This paper has presented a novel MA CF-ISAC system that jointly optimizes AP beamforming and MA positions to enhance sensing accuracy and communication performance under realistic TS errors. By deriving the CRLB in the presence of TS errors and adopting a worst-case optimization approach, we have effectively quantified and mitigate the adverse effects of TS uncertainty. To solve the resulting non-convex optimization problem, we have developed the MA-MetaRL algorithm, aided by an MO method to efficiently adapt beamforming and MA positions across varying environments and constraints. Simulation results have validated that the proposed MA aided CF-ISAC system significantly improves robustness against TS errors, ensures communication quality, and achieves high sensing accuracy, particularly in scenarios with unknown or uncertain TS errors. 

\begin{appendices}
\section{The proof of (\ref{pro14})}\label{APP1}
Based on the receive signal model described in (\ref{pro12}), the corresponding likelihood function can be expressed as follows:
\begin{align}
f(\boldsymbol{y}_{b}|\boldsymbol{d})=\frac{1}{(2\pi(\sigma_{b}^{2})^{S})^{N_{r}/2}}e^{-\frac{1}{2}\frac{(\boldsymbol{y}_{b}-\sum_{a=1}^{A}\boldsymbol{\Xi}_{a,b}\boldsymbol{x}_{a})^{T}\left(\boldsymbol{y}_{b}-\sum_{a=1}^{A}\boldsymbol{\Xi}_{a,b}\boldsymbol{x}_{a}\right)}{\left(\sigma_{b}^{2}\right)^{S}}}.\label{app1}
\end{align}
Thus, the $(i_{1},i_{2})$-th element of the matrix $\boldsymbol{F}_{\boldsymbol{d}\boldsymbol{d}}^{b}$ is given by
\begin{align}
&\boldsymbol{F}_{\boldsymbol{d}\boldsymbol{d}}^{b}(i_{1},i_{2})=-\mathbb{E}\left(\frac{\partial^{2}\ln f(\boldsymbol{y}_{b}|\boldsymbol{d})}{\partial\boldsymbol{d}(i)\partial\boldsymbol{d}(j)}\right)\nonumber\\
&=\frac{2}{\sigma^{2S}}\mathcal{R}\left\{\frac{\partial(\sum_{a=1}^{A}\boldsymbol{\Xi}_{a,b}\boldsymbol{x}_{a})^{H}}{\partial\boldsymbol{d}(i_{1})}\boldsymbol{I}_{b}^{-1}\frac{\partial(\sum_{a=1}^{A}\boldsymbol{\Xi}_{a,b}\boldsymbol{x}_{a})}{\partial\boldsymbol{d}(i_{2})}\right\}\nonumber\\
&+\mathrm{Tr}\left(\boldsymbol{I}_{b}^{-1}\frac{\partial\boldsymbol{I}_{b}}{\partial\boldsymbol{d}(i_{1})}\boldsymbol{I}_{b}^{-1}\frac{\partial\boldsymbol{I}_{b}}{\partial\boldsymbol{d}(i_{2})}\right)=\nonumber\\
&\frac{2}{\sigma^{2S}}\mathcal{R}\left\{\frac{\partial(\sum_{a=1}^{A}\boldsymbol{\Xi}_{a,b}\boldsymbol{x}_{a})^{H}}{\partial\boldsymbol{d}(i_{1})}\frac{\partial(\sum_{a=1}^{A}\boldsymbol{\Xi}_{a,b}\boldsymbol{x}_{a})}{\partial\boldsymbol{d}(i_{2})}\right\},\label{app2}
\end{align}
where $\boldsymbol{d}(i_{1})$ and 
$\boldsymbol{d}(i_{2})$ denote the $i_{1}$-th element of $\boldsymbol{d}$ and the $i_{2}$-th element of $\boldsymbol{d}$, respectively. Based on the expression provided in (\ref{app1}), the derivative of $\sum_{a=1}^{A}\boldsymbol{\Xi}_{a,b}\boldsymbol{x}_{a}$ with respect to the target location parameter $\boldsymbol{d}$ needs to be derived. Accordingly, the derivative of $\sum_{a=1}^{A}\boldsymbol{\Xi}_{a,b}\boldsymbol{x}_{a}$ with respect to the target positions $d_{x}$ and $d_{y}$ is given by
\begin{align}
&\frac{\partial (\sum_{a=1}^{A}\boldsymbol{\Xi}_{a,b}\boldsymbol{x}_{a})}{\partial d_{x}}=\sum_{a=1}^{A}\left(\frac{\partial\boldsymbol{\Xi}_{a,b}}{\partial d_{x}}\boldsymbol{x}_{a}+\underbrace{\boldsymbol{\Xi}_{a,b}\frac{\partial\boldsymbol{x}_{a}}{\partial d_{x}}}_{\boldsymbol{0}}\right)\nonumber\\
&=\sum_{a=1}^{A}\left(\frac{\partial\boldsymbol{\Xi}_{a,b}}{\partial d_{x}}\boldsymbol{x}_{a}\right),\nonumber\\
&\frac{\partial (\sum_{a=1}^{A}\boldsymbol{\Xi}_{a,b}\boldsymbol{x}_{a})}{\partial d_{y}}=\sum_{a=1}^{A}\left(\frac{\partial\boldsymbol{\Xi}_{a,b}}{\partial d_{y}}\boldsymbol{x}_{a}+\underbrace{\boldsymbol{\Xi}_{a,b}\frac{\partial\boldsymbol{x}_{a}}{\partial d_{y}}}_{\boldsymbol{0}}\right)\nonumber\\
&=\sum_{a=1}^{A}\left(\frac{\partial\boldsymbol{\Xi}_{a,b}}{\partial d_{y}}\boldsymbol{x}_{a}\right),\label{app3}
\end{align}
in which 
\begin{align}
&\frac{\partial \boldsymbol{H}_{a,b}}{\partial d_{x}}=-j2\pi f_{s}\beta_{a,b}\frac{\partial\tau_{a,b}}{\partial d_{x}}e^{-j2\pi f_{s}\tau_{a,b}}\boldsymbol{g}_{b}(\boldsymbol{p}_{b})\boldsymbol{g}_{a}^{H}(\boldsymbol{p}_{a})+\beta_{a,b}\nonumber\\
&e^{-j2\pi f_{s}\tau_{a,b}}\frac{\partial \boldsymbol{g}_{b}(\boldsymbol{p}_{b})}{\partial d_{x}}\boldsymbol{g}_{a}^{H}(\boldsymbol{p}_{a})+\beta_{a,b}e^{-j2\pi f_{s}\tau_{a,b}}g_{b}(\boldsymbol{p}_{b})\frac{\partial \boldsymbol{g}_{a}(\boldsymbol{p}_{a})}{\partial d_{x}},\nonumber\\
&\frac{\partial \boldsymbol{H}_{a,b}}{\partial d_{y}}=-j2\pi f_{s}\beta_{a,b}\frac{\partial\tau_{a,b}}{\partial d_{y}}e^{-j2\pi f_{s}\tau_{a,b}}\boldsymbol{g}_{b}(\boldsymbol{p}_{b})\boldsymbol{g}_{a}^{H}(\boldsymbol{p}_{a})+\beta_{a,b}\nonumber\\
&e^{-j2\pi f_{s}\tau_{a,b}}\frac{\partial \boldsymbol{g}_{b}(\boldsymbol{p}_{b})}{\partial d_{y}}\boldsymbol{g}_{a}^{H}(\boldsymbol{p}_{a})+\beta_{a,b}e^{-j2\pi f_{s}\tau_{a,b}}g_{b}(\boldsymbol{p}_{b})\frac{\partial \boldsymbol{g}_{a}(\boldsymbol{p}_{a})}{\partial d_{y}}.\label{app4}
\end{align}
It is evident that to obtain the expression in (\ref{app4}), it is necessary to compute 
$\frac{\partial \boldsymbol{g}_{b}
(\boldsymbol{p}_{b})}{\partial d_{x}}$, 
$\frac{\partial \boldsymbol{g}_{a}^{H}(\boldsymbol{p}_{a})}{\partial d_{x}}$, $\frac{\partial \boldsymbol{g}_{b}
(\boldsymbol{p}_{b})}{\partial d_{y}}$ and 
$\frac{\partial \boldsymbol{g}_{a}^{H}(\boldsymbol{p}_{a})}{\partial d_{y}}$, which are computed as
\begin{small}
\begin{align}
&\frac{\partial \boldsymbol{g}_{b}
(\boldsymbol{p}_{b})}{\partial d_{x}}=\bar{\boldsymbol{g}}_{b}(\boldsymbol{p}_{b})\odot\boldsymbol{g}_{b}(\boldsymbol{p}_{b})\bar{\phi}_{b},\frac{\partial \boldsymbol{g}_{a}^{H}(\boldsymbol{p}_{a})}{\partial d_{x}}=\bar{\boldsymbol{g}}_{a}(\boldsymbol{p}_{a})\odot\boldsymbol{g}_{a}(\boldsymbol{p}_{a})\bar{\phi}_{a},\nonumber\\
&\frac{\partial \boldsymbol{g}_{b}
(\boldsymbol{p}_{b})}{\partial d_{y}}=\bar{\boldsymbol{g}}_{b}(\boldsymbol{p}_{b})\odot\boldsymbol{g}_{b}(\boldsymbol{p}_{b})\tilde{\phi}_{b},\frac{\partial \boldsymbol{g}_{a}^{H}(\boldsymbol{p}_{a})}{\partial d_{y}}=\bar{\boldsymbol{g}}_{a}(\boldsymbol{p}_{a})\odot\boldsymbol{g}_{a}(\boldsymbol{p}_{a})\tilde{\phi}_{a},\label{app5}%
\end{align}
\end{small}%
in which $\bar{\boldsymbol{g}}_{b}(\boldsymbol{p}_{b})$, $\bar{\boldsymbol{g}}_{a}(\boldsymbol{p}_{a})$, $\bar{\tau}_{a,b}$, $\tilde{\tau}_{a,b}$, $\bar{\phi}_{a}$,
$\bar{\phi}_{b}$, $\tilde{\phi}_{a}$ and $\tilde{\phi}_{b}$ are denoted as
\begin{align}
&\bar{\boldsymbol{g}}_{b}(\boldsymbol{p}_{b})=[0,j2\pi p_{1}^{b}\cos(\phi_{b}),\ldots,j2\pi p_{N_{t}}^{b}\cos(\phi_{b})]^{T},\nonumber\\
&\bar{\boldsymbol{g}}_{a}(\boldsymbol{p}_{a})=[0,j2\pi p_{1}^{a}\cos(\phi_{a}),\ldots,j2\pi p_{N_{t}}^{a}\cos(\phi_{a})]^{T},\nonumber\\
&\bar{\tau}_{a,b}=\frac{\partial\tau_{a,b}}{\partial d_{x}}=\frac{1}{c}\frac{(d_{x}^{a}-d_{x})}{\sqrt{(d_{x}^{a}-d_{x})^{2}+(d_{y}^{a}-d_{y})^{2}}}+\nonumber\\
&\frac{1}{c}\frac{(d_{x}^{b}-d_{x})}{\sqrt{(d_{x}^{b}-d_{x})^{2}+(d_{y}^{b}-d_{y})^{2}}},\nonumber\\
&\tilde{\tau}_{a,b}=\frac{\partial\tau_{a,b}}{\partial d_{y}}=\frac{1}{c}\frac{(d_{y}^{a}-d_{y})}{\sqrt{(d_{x}^{a}-d_{x})^{2}+(d_{y}^{a}-d_{y})^{2}}}+\nonumber\\
&\frac{1}{c}\frac{(d_{y}^{b}-d_{y})}{\sqrt{(d_{x}^{b}-d_{x})^{2}+(d_{y}^{b}-d_{y})^{2}}},\nonumber\\
&\bar{\phi}_{a}=\frac{\partial \phi_{a}}{\partial d_{x}}=\frac{(d_{x}^{a}-d_{x})}{(d_{x}^{a}-d_{x})^{2}+(d_{y}^{a}-d_{y})^{2}},\nonumber\\
&\bar{\phi}_{b}=\frac{\partial \phi_{b}}{\partial d_{x}}=\frac{(d_{x}^{b}-d_{x})}{(d_{x}^{b}-d_{x})^{2}+(d_{y}^{b}-d_{y})^{2}},\nonumber\\
&\tilde{\phi}_{a}=\frac{\partial \phi_{a}}{\partial d_{y}}=\frac{(d_{y}^{a}-d_{y})}{(d_{x}^{a}-d_{x})^{2}+(d_{y}^{a}-d_{y})^{2}},\nonumber\\
&\tilde{\phi}_{b}=\frac{\partial \phi_{b}}{\partial d_{y}}=\frac{(d_{y}^{b}-d_{y})}{(d_{x}^{b}-d_{x})^{2}+(d_{y}^{b}-d_{y})^{2}}.\label{app6}
\end{align}
To simplify the analysis, we introduce 
$\bar{\boldsymbol{\Xi}}_{a,b}$ and 
$\tilde{\boldsymbol{\Xi}}_{a,b}$, which are defined as follows:
\begin{align}
&\bar{\boldsymbol{\Xi}}_{a,b}=\frac{\partial \boldsymbol{\Xi}_{a,b}}{\partial d_{x}}=\mathrm{Blkdiag}\left[\frac{\partial \boldsymbol{H}_{a,b,1}}{\partial d_{x}},\ldots,\frac{\partial \boldsymbol{H}_{a,b}}{\partial d_{x}}\right]\nonumber\\
&\tilde{\boldsymbol{\Xi}}_{a,b}=\frac{\partial \boldsymbol{\Xi}_{a,b}}{\partial d_{y}}=\mathrm{Blkdiag}\left[\frac{\partial \boldsymbol{H}_{a,b,1}}{\partial d_{y}},\ldots,\frac{\partial \boldsymbol{H}_{a,b}}{\partial d_{y}}\right].\label{app7}
\end{align}
Therefore, FIM is given at the top of this page.
\begin{figure*}
\begin{align}
\boldsymbol{F}_{\boldsymbol{d}\boldsymbol{d}}^{b}=\left[
\begin{matrix}
\frac{2}{\sigma^{2S}}\mathcal{R}\left\{(\sum_{a=1}^{A}\bar{\boldsymbol{\Xi}}_{a,b}\boldsymbol{x}_{a})^{H}(\sum_{a=1}^{A}\bar{\boldsymbol{\Xi}}_{a,b}\boldsymbol{x}_{a})^{H}\right\}&\frac{2}{\sigma^{2S}}\mathcal{R}\left\{(\sum_{a=1}^{A}\bar{\boldsymbol{\Xi}}_{a,b}\boldsymbol{x}_{a})^{H}(\sum_{a=1}^{A}\tilde{\boldsymbol{\Xi}}_{a,b}\boldsymbol{x}_{a})^{H}\right\}\\
\frac{2}{\sigma^{2S}}\mathcal{R}\left\{(\sum_{a=1}^{A}\tilde{\boldsymbol{\Xi}}_{a,b}\boldsymbol{x}_{a})^{H}(\sum_{a=1}^{A}\bar{\boldsymbol{\Xi}}_{a,b}\boldsymbol{x}_{a})^{H}\right\}&\frac{2}{\sigma^{2S}}\mathcal{R}\left\{(\sum_{a=1}^{A}\tilde{\boldsymbol{\Xi}}_{a,b}\boldsymbol{x}_{a})^{H}(\sum_{a=1}^{A}\tilde{\boldsymbol{\Xi}}_{a,b}\boldsymbol{x}_{a})^{H}\right\}
\end{matrix}
\right].\label{app8}
\end{align}
\hrulefill
\end{figure*}
Based on (\ref{app8}), the $\textbf{CRLB}_{b}(\boldsymbol{d})$ is expressed as
\begin{align}
\mathrm{tr}(\mathrm{\textbf{CRLB}}_{b}(\boldsymbol{d}))=\frac{C}{D}, \label{app9}   
\end{align}
in which $C$ and $D$ are given in (\ref{app10})
\begin{figure*}
\begin{small}
\begin{align}
&A=\frac{2}{\sigma^{2S}}\mathcal{R}\left\{(\sum\nolimits_{a=1}^{A}\bar{\boldsymbol{\Xi}}_{a,b}\boldsymbol{x}_{a})^{H}(\sum\nolimits_{a=1}^{A}\bar{\boldsymbol{\Xi}}_{a,b}\boldsymbol{x}_{a})\right\}+\frac{2}{\sigma^{2S}}\mathcal{R}\left\{(\sum\nolimits_{a=1}^{A}\tilde{\boldsymbol{\Xi}}_{a,b}\boldsymbol{x}_{a})^{H}(\sum\nolimits_{a=1}^{A}\tilde{\boldsymbol{\Xi}}_{a,b}\boldsymbol{x}_{a})\right\}\nonumber\\
&B=\frac{2}{\sigma^{2S}}\mathcal{R}\left\{(\sum\nolimits_{a=1}^{A}\bar{\boldsymbol{\Xi}}_{a,b}\boldsymbol{x}_{a})^{H}(\sum\nolimits_{a=1}^{A}\bar{\boldsymbol{\Xi}}_{a,b}\boldsymbol{x}_{a})\right\}\frac{2}{\sigma^{2S}}\mathcal{R}\left\{(\sum\nolimits_{a=1}^{A}\tilde{\boldsymbol{\Xi}}_{a,b}\boldsymbol{x}_{a})^{H}(\sum\nolimits_{a=1}^{A}\tilde{\boldsymbol{\Xi}}_{a,b}\boldsymbol{x}_{a})\right\}-\nonumber\\
&\frac{2}{\sigma^{2S}}\mathcal{R}\left\{(\sum\nolimits_{a=1}^{A}\bar{\boldsymbol{\Xi}}_{a,b}\boldsymbol{x}_{a})^{H}(\sum\nolimits_{a=1}^{A}\tilde{\boldsymbol{\Xi}}_{a,b}\boldsymbol{x}_{a})\right\}\frac{2}{\sigma^{2S}}\mathcal{R}\left\{(\sum\nolimits_{a=1}^{A}\tilde{\boldsymbol{\Xi}}_{a,b}\boldsymbol{x}_{a})^{H}(\sum\nolimits_{a=1}^{A}\bar{\boldsymbol{\Xi}}_{a,b}\boldsymbol{x}_{a})\right\}.\label{app10}
\end{align}
\end{small}
\hrulefill
\end{figure*}


\section{The proof of (\ref{pro24})}\label{APP2}
According to the chain rule of differentiation, 
$\frac{\partial \mathrm{tr}(\mathrm{\textbf{CRLB}}_{b}(\boldsymbol{d}))}{\partial\boldsymbol{\vartheta}}$ can be expressed as
\begin{align}
\frac{\partial \mathrm{tr}(\mathrm{\textbf{CRLB}}_{b}(\boldsymbol{d}))}{\partial\boldsymbol{\vartheta}}=\frac{\bar{\boldsymbol{C}}D-C\bar{\boldsymbol{D}}}{D^{2}}.\label{app11}
\end{align}
where $\bar{\boldsymbol{C}}$ and $\bar{\boldsymbol{D}}$ are expressed as
\begin{align}
&\bar{\boldsymbol{C}}=\frac{\partial C}{\partial \boldsymbol{\vartheta}}=\frac{2}{\sigma^{2S}}\mathcal{R}\left\{\sum\nolimits_{a^{\prime}=1}^{A}\sum\nolimits_{a=1}^{A}\left(\boldsymbol{x}_{a^{\prime}}^{H}\bar{\bar{\boldsymbol{\Xi}}}\mathrm{vec}(\boldsymbol{X}_{a})(\boldsymbol{I}\otimes\boldsymbol{1})\right.\right.\nonumber\\
&\left.\left.+\boldsymbol{x}_{a^{\prime}}^{H}\bar{\bar{\boldsymbol{\Xi}}}\mathrm{vec}(\boldsymbol{X}_{a})(\boldsymbol{I}\otimes\boldsymbol{1})\right)\right\},\label{app12}
\end{align}
and
\begin{align}
&\bar{\boldsymbol{D}}=\frac{\partial D}{\partial \boldsymbol{\vartheta}}=\frac{2}{\sigma^{2S}}\mathcal{R}\left\{\sum\nolimits_{a^{\prime}=1}^{A}\sum\nolimits_{a=1}^{A}\left(\boldsymbol{x}_{a^{\prime}}^{H}\bar{\bar{\boldsymbol{\Xi}}}\mathrm{vec}(\boldsymbol{X}_{a})(\boldsymbol{I}\otimes\boldsymbol{1})\right)\right\}\nonumber\\
&\times\frac{2}{\sigma^{2S}}\mathcal{R}\left\{(\sum\nolimits_{a=1}^{A}\tilde{\boldsymbol{\Xi}}_{a,b}\boldsymbol{x}_{a})^{H}(\sum\nolimits_{a=1}^{A}\tilde{\boldsymbol{\Xi}}_{a,b}\boldsymbol{x}_{a})\right\}\nonumber\\
&+\frac{2}{\sigma^{2S}}\mathcal{R}\left\{(\sum\nolimits_{a=1}^{A}\bar{\boldsymbol{\Xi}}_{a,b}\boldsymbol{x}_{a})^{H}(\sum\nolimits_{a=1}^{A}\bar{\boldsymbol{\Xi}}_{a,b}\boldsymbol{x}_{a})\right\}\nonumber\\
&\times\frac{2}{\sigma^{2S}}\mathcal{R}\left\{\sum\nolimits_{a^{\prime}=1}^{A}\sum\nolimits_{a=1}^{A}\left(\boldsymbol{x}_{a^{\prime}}^{H}\tilde{\tilde{\boldsymbol{\Xi}}}\mathrm{vec}(\boldsymbol{X}_{a})(\boldsymbol{I}\otimes\boldsymbol{1})\right)\right\}\nonumber\\
&\frac{2}{\sigma^{2S}}\mathcal{R}\left\{\sum\nolimits_{a^{\prime}=1}^{A}\sum\nolimits_{a=1}^{A}\left(\boldsymbol{x}_{a^{\prime}}^{H}\bar{\tilde{\boldsymbol{\Xi}}}\mathrm{vec}(\boldsymbol{X}_{a})(\boldsymbol{I}\otimes\boldsymbol{1})\right)\right\}\nonumber\\
&\times\frac{2}{\sigma^{2S}}\mathcal{R}\left\{(\sum\nolimits_{a=1}^{A}\tilde{\boldsymbol{\Xi}}_{a,b}\boldsymbol{x}_{a})^{H}(\sum\nolimits_{a=1}^{A}\bar{\boldsymbol{\Xi}}_{a,b}\boldsymbol{x}_{a})\right\}\nonumber\\
&+\frac{2}{\sigma^{2S}}\mathcal{R}\left\{(\sum\nolimits_{a=1}^{A}\bar{\boldsymbol{\Xi}}_{a,b}\boldsymbol{x}_{a})^{H}(\sum\nolimits_{a=1}^{A}\tilde{\boldsymbol{\Xi}}_{a,b}\boldsymbol{x}_{a})\right\}\nonumber\\
&\times\frac{2}{\sigma^{2S}}\mathcal{R}\left\{\sum\nolimits_{a^{\prime}=1}^{A}\sum\nolimits_{a=1}^{A}\left(\boldsymbol{x}_{a^{\prime}}^{H}\tilde{\bar{\boldsymbol{\Xi}}}\mathrm{vec}(\boldsymbol{X}_{a})(\boldsymbol{I}\otimes\boldsymbol{1})\right)\right\}.\label{app13}
\end{align}

\end{appendices}


\begin{thebibliography}{99}
\bibitem{10707080} J. Zhang, S. Guo, S. Gong, C. Xing, N. Zhao, D. W. Kwan Ng, and
D. Niyato, “Intelligent waveform design for integrated sensing and
communication,” IEEE Wireless Commun., vol. 32, no. 1, pp. 166–173,
2025.
\bibitem{10666854} W. Lyu, S. Yang, Y. Xiu, X. Chen, Z. Zhang, C. Assi, and C. Yuen,
“Dual-robust integrated sensing and communication: Beamforming un-
der CSI imperfection and location uncertainty,” IEEE Wireless Commun.
Lett., vol. 13, no. 11, pp. 3124–3128, 2024.
\bibitem{11098592} Y. Xiu, Y. Zhao, R. Yang, H. Tang, L. Qu, M. Khabbaz, C. Assi,
and N. Wei, “Latency minimization for movable relay-aided D2D-MEC
communication systems,” IEEE Transactions on Mobile Computing, pp.
1–16, 2025.
\bibitem{9598915} M. Alsabah, M. A. Naser, B. M. Mahmmod, S. H. Abdulhussain, M. R.
Eissa, A. Al-Baidhani, N. K. Noordin, S. M. Sait, K. A. Al-Utaibi, and
F. Hashim, “6G wireless communications networks: A comprehensive
survey,” IEEE Access, vol. 9, pp. 148 191–148 243, 2021.
\bibitem{9598915} X. Dong, W. Lyu, R. Yang, Y. Xiu, W. Mei, and Z. Zhang, “Movable
antenna enhanced secure simultaneous wireless information and power
transfer,” IEEE Communications Letters, pp. 1–1, 2025.
\bibitem{10709885} L. Zhu, W. Ma, Z. Xiao, and R. Zhang, “Performance analysis and
optimization for movable antenna aided wideband communications,”
IEEE Trans. Wireless Commun., vol. 23, no. 12, pp. 18 653–18 668,
2024.
\bibitem{10516289} W. Mao, Y. Lu, C.-Y. Chi, B. Ai, Z. Zhong, and Z. Ding,
“Communication-sensing region for cell-free massive MIMO ISAC
systems,” IEEE Trans. Wireless Commun., vol. 23, no. 9, pp. 12 396–
12 411, 2024.
\bibitem{10494224} Z. Behdad, O. T. Demir, K. W. Sung, E. Bjornson, and C. Cavdar,
“Multi-static target detection and power allocation for integrated sensing
and communication in cell-free massive MIMO,” IEEE Trans. Wireless
Commun., vol. 23, no. 9, pp. 11 580–11 596, 2024.
\bibitem{10605793} Z. Ren, J. Xu, L. Qiu, and D. Wing Kwan Ng, “Secure cell-free
integrated sensing and communication in the presence of information
and sensing eavesdroppers,” IEEE J. Sel. Areas Commun., vol. 42,
no. 11, pp. 3217–3231, 2024.
\bibitem{10742632} Y. Dong, Z. Yang, H. Wang, N. Hao, and H. Li, “Joint user pairing and
beamforming design for NOMA-aided CFMM-ISAC systems,” IEEE
Internet Things J., vol. 12, no. 6, pp. 6749–6763, 2025.
\bibitem{10207026} Y. Cao and Q.-Y. Yu, “Joint resource allocation for user-centric cell-free
integrated sensing and communication systems,” IEEE Commun. Lett.,
vol. 27, no. 9, pp. 2338–2342, 2023.
\bibitem{10540103} A. A. Nasir, “Joint users’ secrecy rate and target’s sensing SNR
maximization for a secure cell-free ISAC system,” IEEE Commun. Lett.,
vol. 28, no. 7, pp. 1549–1553, 2024.
\bibitem{10742291} A. Abdelaziz Salem, M. A. Albreem, K. Alnajjar, S. Abdallah, and
M. Saad, “Integrated cooperative sensing and communication for RIS-
enabled full-duplex cell-free MIMO systems,” IEEE Trans. Commun.,
pp. 1–1, 2024.
\bibitem{10834811} C. Qing, Y. Huang, Q. Zhao, and Q. Hu, “A low-complexity sensing-
aided timing synchronization method for UAV-assisted OFDM systems,”
IEEE Wireless Commun. Lett., vol. 14, no. 3, pp. 911–915, 2025.
\bibitem{10684491} X. Yang, Z. Wei, J. Xu, Y. Fang, H. Wu, and Z. Feng, “Coordinated
transmit beamforming for networked ISAC with imperfect CSI and time
synchronization,” IEEE Trans. Wireless Commun., vol. 23, no. 12, pp.
18 019–18 035, 2024.
\bibitem{10649809} X.-Y. Wang, S. Yang, J. Zhang, C. Masouros, and P. Zhang, “Clutter
suppression, time-frequency synchronization, and sensing parameter
association in asynchronous perceptive vehicular networks,” IEEE J. Sel.
Areas Commun., vol. 42, no. 10, pp. 2719–2736, 2024.
\bibitem{10599127} C. Wang, Z. Li, K.-K. Wong, R. Murch, C.-B. Chae, and S. Jin, “AI-
empowered fluid antenna systems: Opportunities, challenges, and future
directions,” IEEE Wireless Commun., vol. 31, no. 5, pp. 34–41, 2024.
\bibitem{10839251} W. Lyu, S. Yang, Y. Xiu, Z. Zhang, C. Assi, and C. Yuen, “Movable
antenna enabled integrated sensing and communication,” IEEE Trans.
Wireless Commun., vol. 24, no. 4, pp. 2862–2875, 2025.
\bibitem{10696953} H. Qin, W. Chen, Q. Wu, Z. Zhang, Z. Li, and N. Cheng, “Cram´er-Rao
bound minimization for movable antenna-assisted multiuser integrated
sensing and communications,” IEEE Wireless Commun. Lett., vol. 13,
no. 12, pp. 3404–3408, 2024.
\bibitem{10870338} Y. Xiu, S. Yang, W. Lyu, P. L. Yeoh, Y. Li, and Y. Ai, “Movable antenna
enabled ISAC beamforming design for low-altitude airborne vehicles,”
IEEE Wireless Commun. Lett., pp. 1–1, 2025.
\bibitem{10962171} H. Wu, H. Ren, C. Pan, and Y. Zhang, “Movable antenna-enabled
RIS-aided integrated sensing and communication,” IEEE Trans. Cognit.
Commun. Networking, pp. 1–1, 2025.
\bibitem{10693833} Z. Kuang, W. Liu, C. Wang, Z. Jin, J. Ren, X. Zhang, and Y. Shen,
“Movable-antenna array empowered ISAC systems for low-altitude
economy,” in 2024 IEEE/CIC International Conference on Communi-
cations in China (ICCC Workshops), 2024, pp. 776–781.
\bibitem{10901248} A. Khalili and R. Schober, “Advanced ISAC design: Movable antennas
and accounting for dynamic RCS,” in GLOBECOM 2024 - 2024 IEEE
Global Communications Conference, 2024, pp. 4022–4027.
\bibitem{1599595} Y. Mostofi and D. Cox, “Mathematical analysis of the impact of timing
synchronization errors on the performance of an OFDM system,” IEEE
Trans. Commun., vol. 54, no. 2, pp. 226–230, 2006.
\bibitem{10243545} W. Ma, L. Zhu, and R. Zhang, “MIMO capacity characterization for
movable antenna systems,” IEEE Trans. Wireless Commun., vol. 23,
no. 4, pp. 3392–3407, Sep. 2024.
\bibitem{10032173} Y. Ju, H. Wang, Y. Chen, T.-X. Zheng, Q. Pei, J. Yuan, and N. Al-
Dhahir, “Deep reinforcement learning based joint beam allocation and
relay selection in mmWave vehicular networks,” IEEE Trans. Commun.,
vol. 71, no. 4, pp. 1997–2012, 2023.
\bibitem{7961152} K. Venugopal, A. Alkhateeb, N. Gonz´alez Prelcic, and R. W. Heath,
“Channel estimation for hybrid architecture-based wideband millimeter
wave systems,” IEEE J. Sel. Areas Commun., vol. 35, no. 9, pp. 1996–
2009, 2017.
\bibitem{ning2025movable} B. Ning, S. Yang, Y. Wu, P. Wang, W. Mei, C. Yuen, and E. Bjornson,
“Movable antenna-enhanced wireless communications: General archi-
tectures and implementation methods,” IEEE Wireless Commun., 2025.
\bibitem{6616600} K. Xu, Y. Xu, W. Ma, W. Xie, and D. Zhang, “Time and frequency syn-
chronization for multicarrier transmission on hexagonal time-frequency
lattice,” IEEE Trans. Signal Process., vol. 61, no. 24, pp. 6204–6219,
2013.
\bibitem{4400801} Y. Mostofi and D. C. Cox, “A robust timing synchronization design
in ofdm systems-part i: low-mobility cases,” IEEE Trans. Wireless
Commun., vol. 6, no. 12, pp. 4329–4339, 2007.
\bibitem{10477314} C. Wang, G. Li, H. Zhang, K.-K. Wong, Z. Li, D. W. K. Ng, and C.-B.
Chae, “Fluid antenna system liberating multiuser MIMO for ISAC via
deep reinforcement learning,” IEEE Trans. Wireless Commun., vol. 23,
no. 9, pp. 10 879–10 894, 2024.
\bibitem{9472958} Y. Xiu, J. Zhao, W. Sun, M. D. Renzo, G. Gui, Z. Zhang, and N. Wei,
“Reconfigurable intelligent surfaces aided mmWave NOMA: Joint power
allocation, phase shifts, and hybrid beamforming optimization,” IEEE
Trans. Wireless Commun., vol. 20, no. 12, pp. 8393–8409, Jul. 2021.
\bibitem{10797657} Y. Xiu, Y. Zhao, S. Yang, Y. Zhang, D. Niyato, H. Du, and N. Wei,
“Robust beamforming design for near-field DMA-NOMA mmwave
communications with imperfect position information,” IEEE Trans.
Wireless Commun., vol. 24, no. 2, pp. 1678–1692, 2025.
\bibitem{absil2010optimization} P.-A. Absil, R. Mahony, and R. Sepulchre, “Optimization on manifolds:
Methods and applications,” in Recent Advances in Optimization and its
Applications in Engineering: The 14th Belgian-French-German Confer-
ence on Optimization. Springer, 2010, pp. 125–144.
\bibitem{gb08} M. Grant and S. Boyd, “Graph implementations for nonsmooth convex
programs,” in Recent Advances in Learning and Control, ser. Lecture
Notes in Control and Information Sciences, V. Blondel, S. Boyd, and
H. Kimura, Eds. Springer-Verlag Limited, 2008, pp. 95–110.
\bibitem{fujimoto2018addressing} S. Fujimoto, H. Hoof, and D. Meger, “Addressing function approxi-
mation error in actor-critic methods,” in International conference on
machine learning. PMLR, 2018, pp. 1587–1596.
\bibitem{10480601} Y. Gao, X. Yuan, D. Yang, Y. Hu, Y. Cao, and A. Schmeink, “UAV-
assisted MEC system with mobile ground terminals: DRL-based joint
terminal scheduling and UAV 3D trajectory design,” IEEE Trans. Veh.
Technol., pp. 1–17, Mar. 2024.
\end{thebibliography}
\end{document}